\documentclass{article}
\usepackage[numbers]{natbib}
\usepackage{graphicx}
\usepackage{tabularx}
\usepackage{amsmath}
\usepackage{float}
\usepackage{booktabs}
\usepackage[left]{lineno} % line numbers
\usepackage{authblk} % for proper author affiliations display
\usepackage{threeparttable} % for notes under tables. No idea why that's not part of the default functionality.

\usepackage[printwatermark]{xwatermark}
\usepackage{xcolor}
\newwatermark[allpages,color=red!30,angle=90,scale=1,xpos=91,ypos=0]{Preprint -- Under peer review}

\usepackage[font=bf]{caption} % Makes figure caption bold
\renewcommand\figurename{Fig.} % Abbreviates Figure in caption
\usepackage[top=2.5cm,right=2.5cm, left=2.5cm, bottom=2.5cm]{geometry} % page margins
\usepackage{titlesec}
\titleformat*{\subsection}{\normalsize\bfseries}

\usepackage{hyperref}
\hypersetup{
    colorlinks=true,
    citecolor=black, 
    linkcolor=black,
    filecolor=black,      
    urlcolor=blue,
    }
\usepackage{cleveref}

\begin{document}
\widowpenalty10000 % to prevent single/ half lines on new pages
%\linenumbers 
% textwidth in cm: \printinunitsof{cm}\prntlen{\textwidth} 16.04 cm
% Title and author details. 
\title{An installation-level model of China's coal sector shows how its decarbonization and energy security plans will reduce overseas coal imports}
\author[1,*]{Jorrit Gosens}
\author[2]{Alex Turnbull}
\author[1]{Frank Jotzo}
\affil[1]{\footnotesize Crawford School of Public Policy, Australian National University, Acton, Australian Capital Territory, Australia}
\affil[2]{\footnotesize Keshik Capital, Singapore}
\affil[*]{\footnotesize Corresponding author; jorrit.gosens@anu.edu.au}
\date{} % exclude date below title

\maketitle
\begin{abstract}
China aims for net-zero carbon emissions by 2060, and an emissions peak before 2030. This will reduce its consumption of coal for power generation and steel making. Simultaneously, China aims for improved energy security, primarily with expanded domestic coal production and transport infrastructure.

Here, we analyze effects of both these pressures on seaborne coal imports, with a purpose-built model of China's coal production, transport, and consumption system with installation-level geospatial and technical detail. This represents a 1000-fold increase in granularity versus earlier models, allowing representation of aspects that have previously been obscured.

We find that reduced Chinese coal consumption affects seaborne imports much more strongly than domestic supply. Recent expansions of rail and port capacity, which reduce costs of getting domestic coal to Southern coastal provinces, will further reduce demand for seaborne thermal coal and amplify the effect of decarbonisation on coal imports. Seaborne coking coal imports are also likely to fall, because of expanded supply of cheap and high quality coking coal from neighbouring Mongolia.
\end{abstract}

\section*{}
\begin{figure}[t] % hbtp for here, top, bottom, next available page. H for really just here
    \centering
    \includegraphics[width=1.0\textwidth]{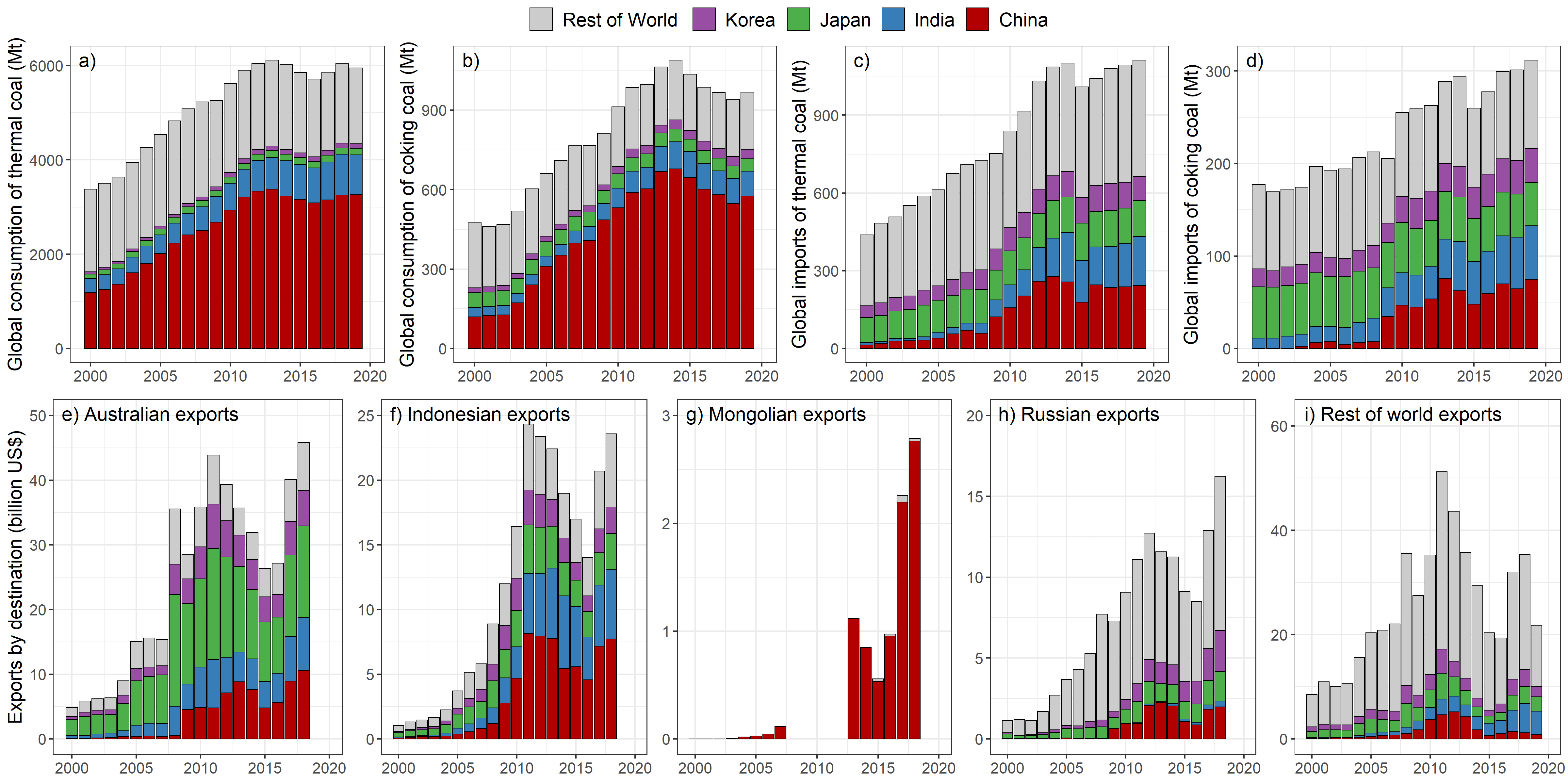}
        \caption{Global coal markets. 
        \textmd{Global consumption of thermal (a) and coking coal (b); global imports of thermal (c) and coking coal (d); and revenue by destination for key exporters of coal (e-i; total of coking and thermal coal combined). Data source: (a-d): IEA\cite{IEA/OECD2019}, (e-i): UN Comtrade \cite{UNTradeStatisticsBranch2019}, panel (f) adjusted on the basis of Chinese customs data \cite{ShanxiFenshengInformationServicesCo.Ltd.2021SXcoalData}}.
        }
    \label{fig:coal_cons}
\end{figure}

Reducing the consumption of coal, used primarily in power generation and steelmaking, is a key element of net-zero emissions plans or other long-term low-emissions strategies. This will affect the market outlook for coal exporting countries, in particular as countries with domestic coal mining industries may seek to limit negative effects on domestic coal mining industries \cite{Harrahill2019FrameworkJurisdictions}. China's decarbonization plans are particularly relevant in this respect, as it is the world's largest consumer of both thermal and coking coal, and a large source of revenue for key exporters in the region, in particular Australia and Indonesia (Fig.~\ref{fig:coal_cons}).

Global coal markets have been the subject of much earlier research, and China has been a focal country in most such studies given its dominant share in global coal production, consumption and imports. Yet, exactly how Chinese coal imports depend on domestic market and other developments remains poorly assessed.

In particular, such previous studies have typically used linear cost optimization with so-called multi-regional models, or node-and-link type models, to represent transport infrastructure. It’s important to accurately account for transport in such analyses, given the large share of transport in the total cost of coal to consumers, and because of the restrictions imposed by technical transport capacities of this infrastructure such as railways and ports. However, previous work for such analyses has always used highly simplified networks, with even the most granular models using a few dozen nodes representing continents in global analysis, or provinces in China-focused analyses. These nodes conflate provincial-level production and demand into single points, and inter-provincial transport infrastructure into single links. This casts doubt on just how accurately transport costs and capacity limits are really considered, and therefore how accurately the relative competitiveness of local and imported coal are assessed, in such models.

We drastically improve on this state of the art, with an installation-level model of China, that is, a model that represents every coal mine, power and steel plant, all ports, all railways with all stops, and an intercity road network. Our model has 12,000 nodes and 40,000 links with accurate technical and geo-spatial detail for individual installations. Calibration of this model versus real data for the year 2015-2019 shows its very high accuracy, giving confidence about its predictions for future years. 

We use this model to assess the sensitivity of levels of imported coal versus changes in Chinese consumption of thermal and coking coal. We also consider different future scenarios as to where in China changes in coal consumption would happen, and assess the likely effects of such geographic shifts in consumption on coal imports. Lastly, we consider different counterfactual scenarios, where we assess what import levels would have been in the absence of recent infrastructural investments. 

We find that overseas imports of thermal and coking coal coal are expected to fall between 52 and 96 Mt by 2025 and between 56 and 124 Mt by 2030, from 209 Mt in 2019, depending on how rapidly, and where in China, decarbonization occurs. For coking coal, overseas imports are expected to fall between 9 and 13 Mt by 2025 and between 12 and 17 Mt by 2030, from 68 Mt in 2019, getting pushed out mostly by increased overland imports of cheap high quality coking coal from neighbouring Mongolia. If China had not made its recent substantial investments in its coal transport infrastructure, imports in 2025 and 2030 would have remained at roughly similar levels to 2019 in high future demand scenarios, and would have fallen substantially less in low future demand scenarios.

\section*{Factors determining China's demand for coal imports}
Although China imported only 7.2\% of the thermal coal and 13.0\% of the coking coal it consumed in 2019 \cite{IEA/OECD2019}, it is the world's largest importer by total volume (Fig.~\ref{fig:coal_cons}). 

This Chinese reliance on imported coal cannot be explained by a limited domestic resource availability. Its total proved reserves stood at 141.6 Gt by the end of 2019, behind only the U.S., Russia, and Australia \cite{BP2020Statistical2019}. Whilst production capacity had a difficult time keeping in step with rapidly growing consumption in the 2000's, a growth of investment in coal mining capacity at 32\% p.a. over 2000 to 2012 and slowing consumption growth since then have largely balanced out again \cite{Wang2019ProvincialFairness}. Currently, reserves of operational mines were approximately 94.6 Gt, or about 24 years worth at 2019 consumption levels \cite{WoodMackenzie2021China2021}.

The geological conditions in China are also not such that production costs are exceptionally high \cite{Haftendorn2012TheModel}. Chinese mines can produce 3,200 Mt of thermal coal per year, roughly the amount consumed in China annually, at a mine-gate cost (that is, excluding transport) of \$60/t or less; comparable to or below usual prices in global export markets \cite{WoodMackenzie2021China2021}.

\begin{figure}[t] % hbtp for here, top, bottom, next available page. H for really just here
    \centering
    \includegraphics[width=1\textwidth]{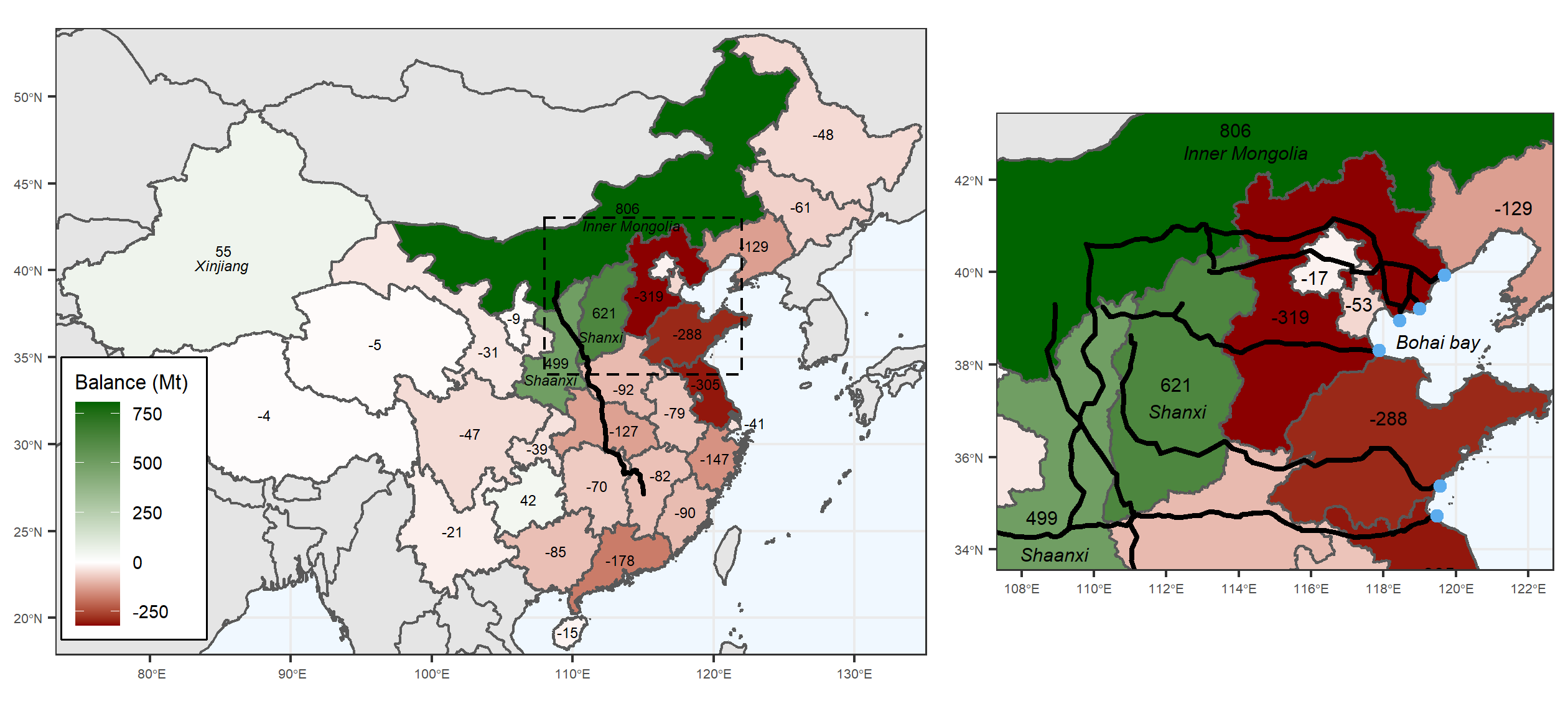}
        \caption{Provincial-level coal supply balance in China. 
        \textmd{Supply balance as provincial level production - consumption. The full map on the left shows the route of the recently opened HaoJi railway. The cutout on the right shows major coal rail lines (black lines) and coal handling ports (blue dots). Consumption and production data source: SXcoal \cite{ShanxiFenshengInformationServicesCo.Ltd.2021SXcoalData}.}
        }
    \label{fig:coal_balance}
\end{figure}

There is geographical imbalance in Chinese coal production and consumption. China's top 4 coal producers -Inner Mongolia, Shaanxi, Shanxi, and Xinjiang- produce about 79\% of all domestic output. These are inland, Northern provinces, whilst coal consumption is concentrated in coastal provinces (Fig.~\ref{fig:coal_balance}). China has a number of high capacity rail lines dedicated to the transport of coal, almost all of which start in these coal basins and heads East towards the industrial heartland around Bohai Bay (Fig.~\ref{fig:coal_balance}), where China's heavy industry including its steel making industry is concentrated. This is also where these rail lines connect with China's major coal ports, where coal is shipped South, in large part for electricity generation in coastal power plants. These coastal power plants have their own port facilities, which also allow international imports. 

There is government regulation in almost every aspect of the coal sector, including mining permits, pricing, and limits on utilization of railways for coal, all of which impact competitiveness of foreign coals \cite{Shi2018UnintendedPolicy, Yang2018UnifyingEnterprises}. Further, although never disclosed in official documents as far as we are aware, China is also considered to use import quota, which are relaxed in times of temporary supply shortages \cite{SPGlobalPlatts2020ChinaSources}. 

\section*{China's plans for emissions reductions and energy security}
China is responsible for about 28\% of global carbon dioxide emissions from energy \cite{TheGlobalCarbonProjectGCP2021GlobalAtlas}, driven by its heavy reliance on coal for power generation (4,631 TWh or 53\% of all global coal-fired power in 2020 \cite{EmberClimate2020GlobalReview}), and its very high steel output (1,053 Mt or 56\% of global output in 2020 \cite{WorldSteelAssociation2021World2020}). 

However, China is also rapidly building out its renewable electricity generation capacity, with wind and solar generating 9.5\% of all power by 2020, up from 1.2\% in 2010 \cite{ChinaElectricityCouncil2020Basic2019}. Steel making, too, has become less carbon-intensive over time, due to increasing availability and use of steel scrap \cite{Xuan2016ForecastChina}. This replaces some of the demand for primary steel, made with iron ore and coking coal in conventional blast oxygen furnace steel making. Continued cost reductions of renewables and further growth in the availability of scrap in China can be expected. China's recently announced goal of net-zero carbon emission by 2060, with an interim goal of peak carbon emissions by 2030, are likely to further accelerate these decarbonization processes, although the policies that will determine the emissions trajectory are still being designed \cite{Dong2021ChinasPaths}. The inevitably reduced consumption of coal can be expected to affect coal imports.

In an effort to improve energy security and use more of the domestically mined coal, China has also invested heavily in transport capacity. Between 2011 and 2020, over a period where total coal consumption was relatively stable (Fig.~\ref{fig:coal_cons}), the network length for freight rail lines used for the transport of coal grew from circa 36,000 to 46,600 km, with annual transport capacity of these lines growing from 1,788 Gt$\cdot$km to 4,691 Gt$\cdot$km. This includes the 1,800 km HaoJi rail line, which started service in 2021. This line has a capacity of about 200 Mt per year, and is dedicated to the transport of coal, from the Ordos basin in Inner Mongolia to Central China's inland provinces (Fig.~\ref{fig:coal_balance}). The handling capacity of the major coal ports in the North of China grew from 760 to 900 Mt per year over the same period \cite{WoodMackenzie2021China2021, Rioux2016}. 

China is further expanding rail links into neighboring Mongolia, which is rapidly opening up mining capacity for both thermal and high quality coking coals. 

Lastly, China is building the largest ultra-high voltage (UHV) transmission network in the world. In some cases these UHV lines allow better use of the large hydropower, wind or solar PV bases in China's West. In other cases they transmit electricity from multi-GW coal-fired power plants in coal mining regions towards demand centers in the east, displacing some of the power generation in those coastal provinces, which import most coal \cite{Paulus2011CoalMarket,TheLantauGroup2020CuttingEvolves}. 

\section*{Previous China coal supply models and assessments}
A common approach in studies assessing international coal market developments is with multi-regional coal supply models, which are linear optimization models that seek to satisfy a given level of demand for coal whilst minimizing the total cost of production and transport of coal. At the core of such models are node \& link networks that represent transport infrastructure between producing and consuming regions. Proper representations of transport infrastructure capacities and transport costs is key because of the significant share of transport in total costs, and because of the large differences in costs of transport via truck, rail, or ship. 

Such multi-regional coal supply models have been used, for example, to investigate which coal-producing regions would optimally supply different consumers,  in different scenarios of future growth or decline of global coal consumption, and/or what coal prices would be in different demand and supply scenarios \cite{Haftendorn2012TheModel, Paulus2011CoalMarket, Truby2012MarketTrade, Ansari2020Between2055, Auger2021TheAssets}. Both Haftendorn et al \cite{Haftendorn2012TheModel} and Paulus and Truby \cite{Paulus2011CoalMarket} forecast that rising Chinese consumption would lead to increases in Chinese imports exceeding 200 Mt between 2010 and 2030, under business as usual scenarios. Building on a more recent BAU scenario developed by the IEA, Auger et al \cite{Auger2021TheAssets} forecast a circa 27 Mt drop in Chinese imports, under essentially stable levels of consumption through 2030.

The same type of models have been used in a string of China-focused analyses. A number of these analyses focused on substitution between different types of energy, and concluded there will be strong downward pressure on coal demand and prices from competition with renewables or other fuels, though with little explicit assessment on volumes of coal imported \cite{Jie2021TheStrategies, Zhang2018ASystem, Li2019TheMarket}. Rioux et al \cite{Rioux2016} calculated that most of China's government plans for investments in rail and port infrastructure in the early 2010s would have highly positive economic benefits. They estimated that bottle-necks in the coal transport system increased total system supply costs by circa \$35 billion, largely because the high transport costs made domestic coals uncompetitive versus imported coals, in particular for consumers in China's Southern coastal provinces. A similar analysis focused on the potential expansion of UHV transmission found that this could lead to \$149 billion in net welfare gains by 2030 \cite{Paulus2011CoalMarket}. Both papers estimated that the planned infrastructure investments might crowd out imports of coal entirely \cite{Rioux2016, Paulus2011CoalMarket}. Other analyses assessed China's policies for cutting over-capacity coal mining from 2016 onwards, which were a bid to tackle falling profitability in the coal mining sector. The policy was found to be so effective that prices rose almost 50\%, strongly increasing competitiveness of imported coal \cite{Shi2018UnintendedPolicy, Wang2019ProvincialFairness, Wu2020ReducingApproach}.

Whilst these analyses have provided valuable insights into different and specific aspects of the economics of the global or Chinese coal supply system, their utility in answering how Chinese imports depend on domestic production capacity, transport capacities, and consumption levels, may still be limited, due to the limited geo-spatial detail in constructing these multi-regional models. Specifically, the regions, or nodes, in these models typically aggregate large geographic areas, with China represented as about 5 or 6 regional nodes in global models, whilst even the most granular models described above use Chinese provinces as individual nodes. 

Such nodes aggregate all supply and demand for coal, as well as all port loading capacity within the province(s) they cover, whilst links between nodes aggregate all rail capacity and road connections between them. This effectively gives the appearance that each consumer within the province(s) is equally well connected to all transport networks. The reality, as also identified in a plant-by-plant review of satellite images for this research (see Methods), is that a large share of steel and power plants are directly on the coast or along rivers, and have port facilities but not railway connections. They are therefore very poorly connected to the landborne market, and almost entirely dependent on seaborne supplies, whether imported or shipped from northern ports. 

A related aspect is the recognition that so many of these plants along the ocean or rivers have their own port facilities for ocean-going vessels or river barges. None of the papers referred to above specifies what values were used for port capacities of different nodes, so we cannot determine if these were included or not. However, none of them mentions specifically these port facilities at power and steel plants, and some refer to capacities for `major' or `main' ports only \cite{Rioux2016, Shi2018UnintendedPolicy, Yang2018UnifyingEnterprises}. These are highly relevant to the seaborne market: in Jiangsu, the largest consumer of coal along the southern coast, total port capacity in 2020 was 593 Mt/yr, of which 273 Mt/yr located at power or steel plants with their own port facilities, in Guangdong, the second largest southern consumer, it was 137 Mt out of a total of 407 Mt of annual unloading capacity. 

Lastly, the aggregation of railways between a limited number of nodes creates the false impression that e.g., all East-West bound lines are perfectly interconnected with all North-South lines, and that all rail lines within a certain area are connected with all port capacity in the same area. This aggregation may have been a sensible simplification in earlier research, but with the current availability of (open source) information on power plants, steel plants, and railway networks, it is not a necessary simplification anymore.

\section*{An installation-level model of China's coal market}
Here, we introduce a multi-regional model of China's coal production, transportation, and consumption system, with installation-level geospatial and technical detail. The `XX China Coal model', or `XX-CCM' (name censored for blind review), includes a network with tens of thousands of nodes and links (Table \ref{tab:nodes-links}); roughly a thousand-fold improvement in granularity versus the current the state of the art, allowing representation of aspects that have been obscured in earlier models.

Each individual coal-fired power plant unit and steel plant is represented as a separate demand node. For each of these plants we have precise location, as well as technical characteristics such as generation or production capacity, and approximate conversion efficiency.

\begin{figure}[H] % hbtp for here, top, bottom, next available page. H for really just here
    \centering
    \includegraphics[width=1\textwidth]{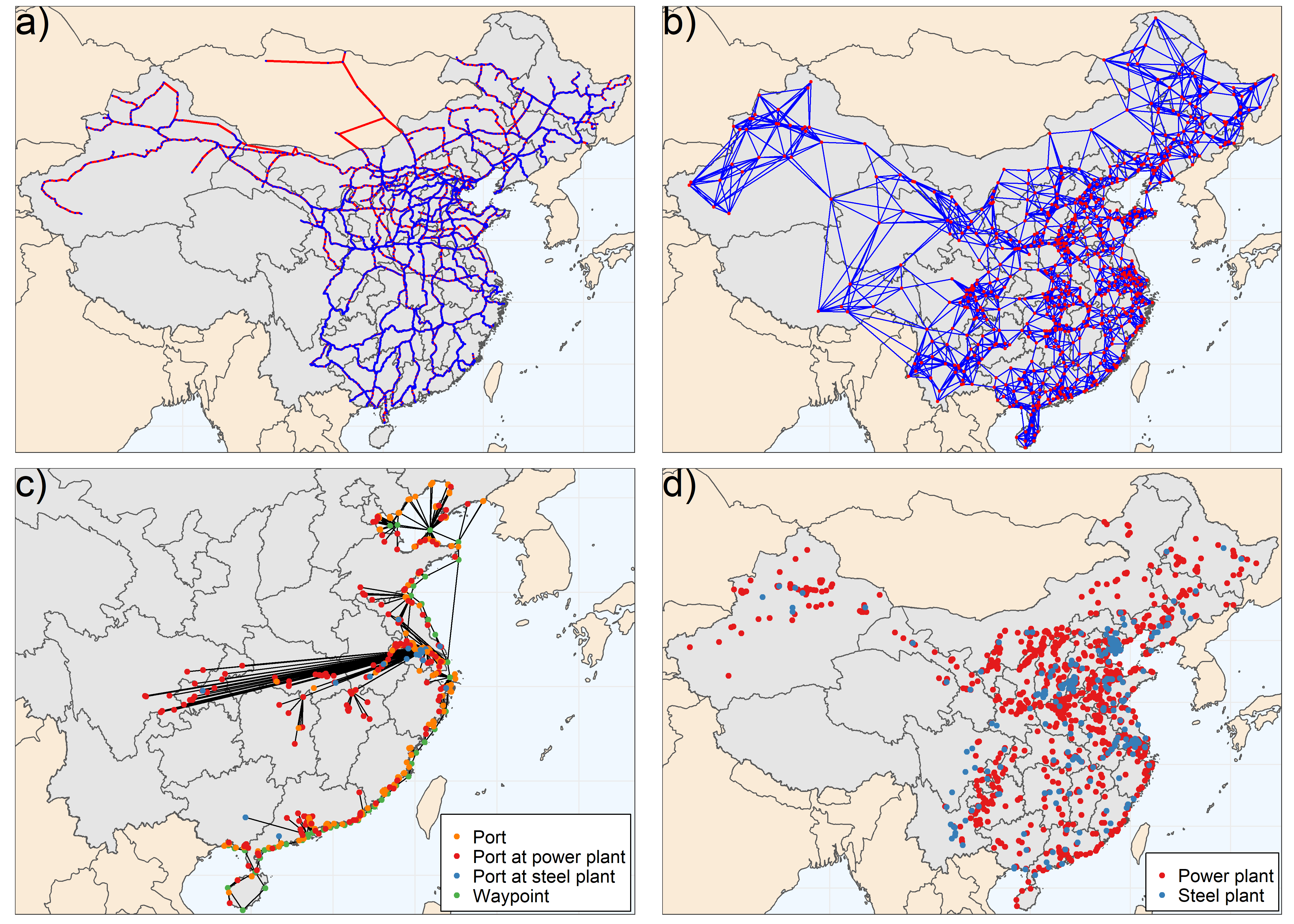}
        \caption{Main transport networks in our model.
        \textmd{a) Railway network; includes only freight railway lines that are used for the transport of coal; railway stops in blue, railway links in red: b) inter-city road transport network; city nodes in red, road links in blue; note the links are drawn here as straight lines but actual driving distances have been used to calculate transport costs over each link; c) navigation network, including ports and navigational way-points; d) power plants and steel plants. Larger scale maps are provided in Supplementary Fig. \ref{fig:rail-network}-\ref{fig:power-steel-plants-map}.}
        }
    \label{fig:maps}
\end{figure}

The model also includes a highly granular representation of China's coal transport networks. First, it includes a rail network, with route, distances between railway stops, and transport capacity of 145 freight rail lines. Second, a road network with links between all of China's 685 cities, and between other relevant types of nodes, with actual driving distances determined for each link. Third, a navigation network with precise location and handling capacity for 483 individual ports, including port facilities at coastal steel and power plants. Fourth, a representation of China's rapidly expanding Ultra-High Voltage (UHV) network for the long-distance transmission of electricity. The network includes accurate interconnections between different rail lines, connections to the rail network for coastal power plants and all steel plants, and mine-mouth operations (direct connections between mines and power plants) in key coal-bearing areas. An overview of node and link types is provided in (Table \ref{tab:nodes-links}), with illustrative maps of the key transport networks and locations of power and steel plants in Fig. \ref{fig:maps}. 

This network is the main component in a linear optimization model that minimizes total production and transport costs, at exogenously determined levels of demand for thermal and coking coal.

\begin{table}[H] % hbtp for here, top, bottom, next available page. H for really just here
\caption{Components of the transport network in our model}
\label{tab:nodes-links}
\begin{threeparttable}
\begin{tabularx}{\textwidth}{lllll}
\toprule
Node type                        & Number           & Link type            & Number &  \\ 
\midrule
Power plants / units             & 1,680 / 3,364    & Railway              & 9.938   &  \\
Steel plants                     & 276              & Inter-city road network & 6,132   &  \\
Railway stops                    & 4,743            & Other road connections & 22,790   &  \\
City centers                     & 685              & River or ocean        & 1,431    &  \\ 
Navigation waypoints             & 103              & UHV network           & 62 &  \\
Ports                            & 483              & Minemouth connections & 43 &  \\
Mines (or mine clusters) domestic / foreign & 174 / 542   & & &\\ 
Prov. level power / steel demand centers  & 30 / 28   &                     &       &  \\ 
\midrule
Total nodes                      &  12,161          &  Total links & 40,396 &  \\ 
\midrule
\end{tabularx}
    \begin{tablenotes}
      \small
      \item Note: total nodes includes a small number of node types not specified here. Further details in the Methods section.
    \end{tablenotes}
\end{threeparttable}
\end{table}

\section*{Sensitivity of Chinese coal imports to consumption levels}
We use our model to determine which mines, domestic or foreign, would supply thermal and coking coal, at various levels of Chinese demand. We purposely use a very wide range of future growth or reduction of consumption, in order to gauge the relationship between Chinese coal consumption levels and imported volumes of coal. 

The baseline model setting includes all transport infrastructure that is either operational or well under construction, and distributes the given future consumption of thermal and coking coal within China on the basis of historical patterns (more details the Methods section).

\begin{figure}[t] % hbtp for here, top, bottom, next available page. H for really just here
    \centering
        \includegraphics[width=1\textwidth]{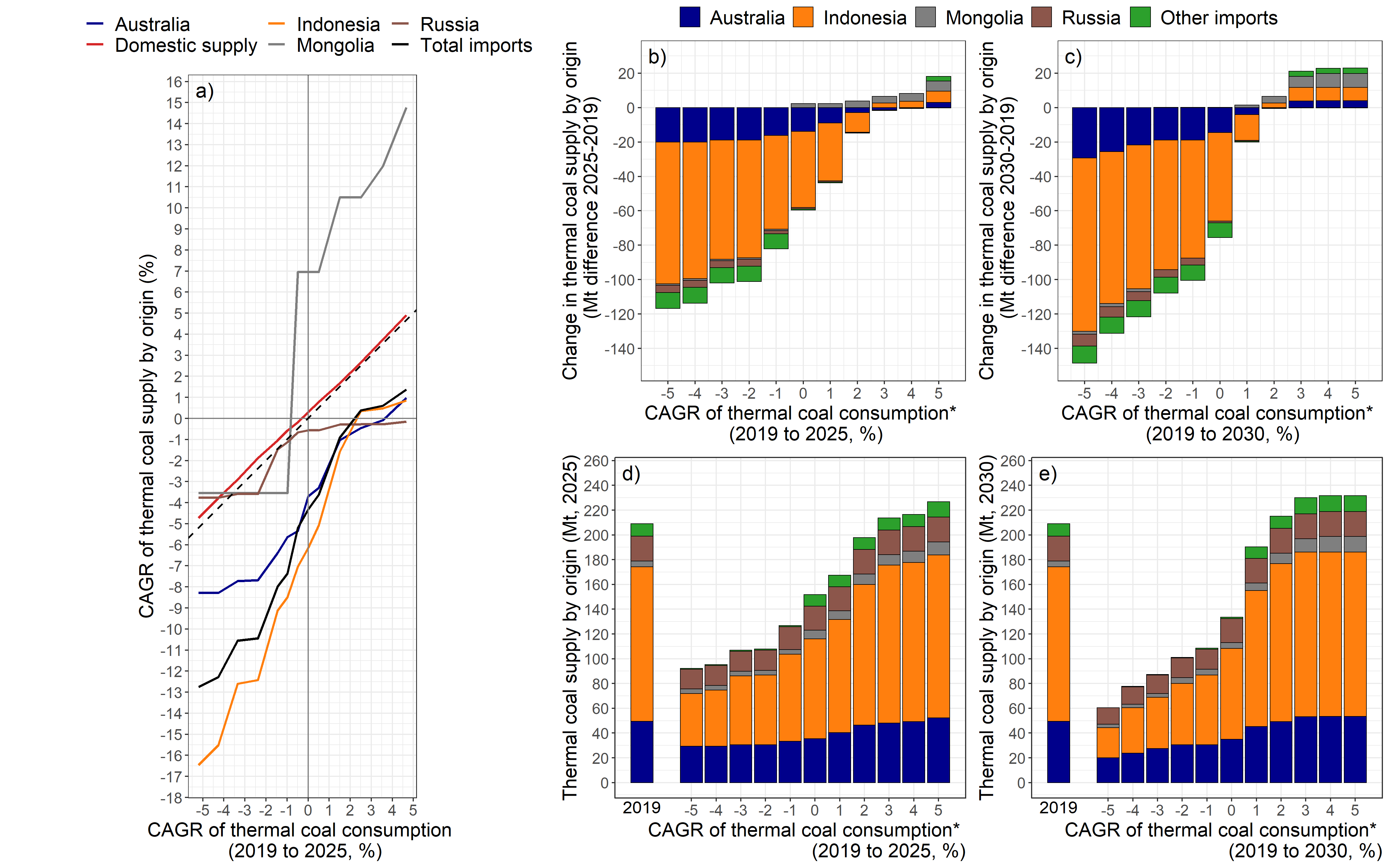} 
    \caption{Sensitivity of Chinese imports of thermal coal to changes in domestic consumption. 
    \textmd{As compound annual growth rates (CAGR, in \%) of imports vs. domestic consumption (a); as Mt changes through 2025 (b) and 2030 (c); and as resulting Mt of total supply in 2025 (d) and 2030 (e). The dotted line in panel a) is drawn at x=y and represents total domestic thermal coal consumption. The left most bars in (d) and (e) represent 2019 levels (as modelled, see the method section on model calibration). The CAGR in panels (b-e) are actually the CAGR of coal-fired power production and the consumption of thermal coal in other industries. The CAGR of thermal coal differs slightly from that number due to improvements in power plant efficiency. Presentation of this sensitivity in panel (a) as total percentage change and absolute changes in Mt through 2025 is included in Supplementary Fig. \ref{fig:sensitivity_thermal_alt}.}
    }
    \label{fig:sensitivity_thermal}
\end{figure}

\begin{figure}[t] % hbtp for here, top, bottom, next available page. H for really just here
    \centering
        \includegraphics[width=1\textwidth]{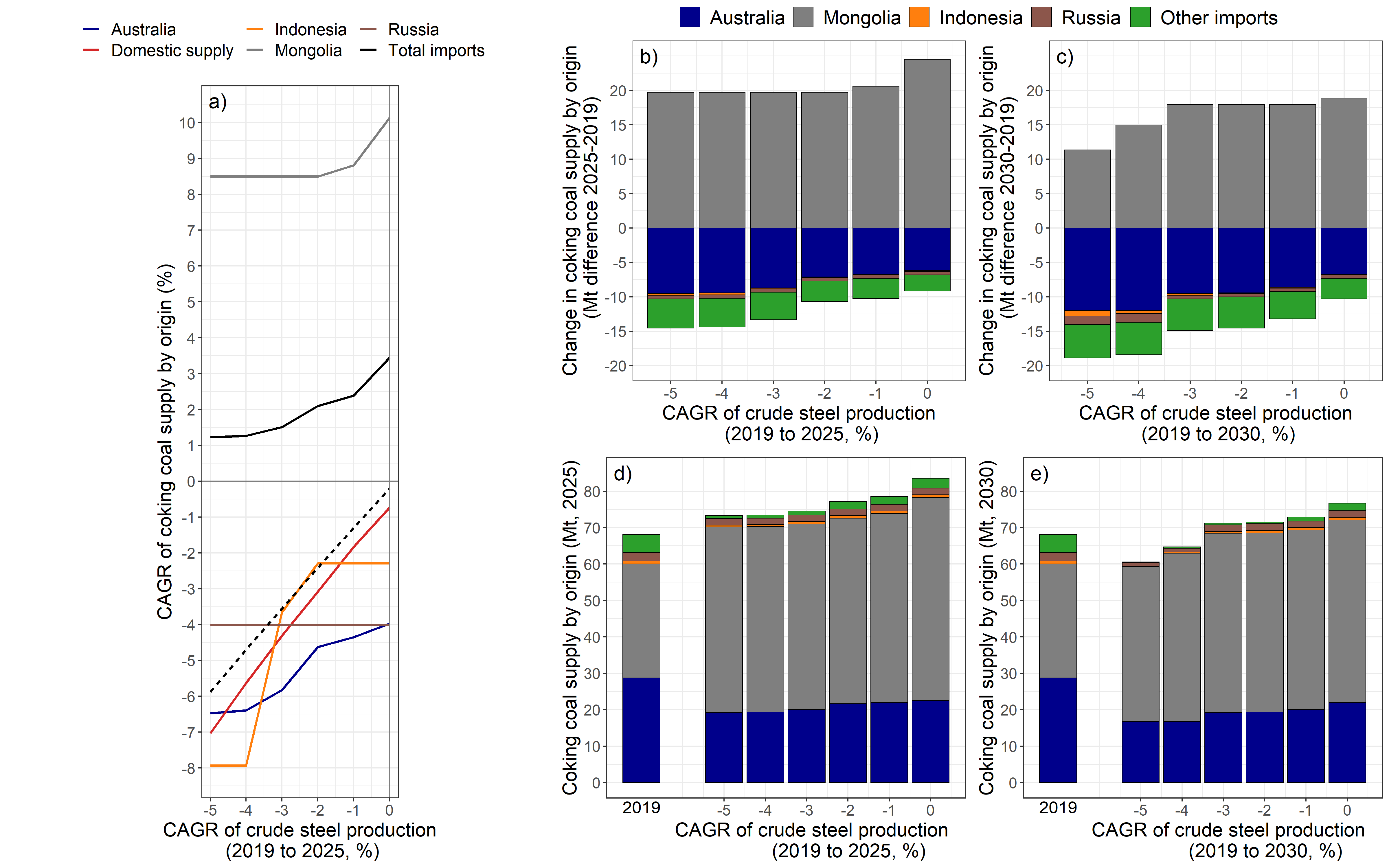} 
    \caption{Sensitivity of Chinese imports of coking coal to changes in domestic consumption. 
        \textmd{As compound annual growth rates (CAGR, in \%) of imports vs. domestic consumption (a); as Mt changes through 2025 (b) and 2030 (c); and as resulting Mt of total supply in 2025 (d) and 2030 (e). The dotted line in panel a) represents total domestic coking coal consumption; that line is just below the line x=y, as a share of steel production is done with recycled scrap steel rather than iron ore and coking coal. The left most bars in (d) and (e) represent 2019 levels (as modelled, see the method section on model calibration). Presentation of this sensitivity in panel (a) as total percentage change and absolute changes in Mt through 2025 is included in Supplementary Fig. \ref{fig:sensitivity_coking_alt}.}
    }
    \label{fig:sensitivity_coking}
\end{figure}

Results suggest that domestic coal production will closely track any growth or reduction in domestic coal consumption to 2025 and 2030 (Fig. \ref{fig:sensitivity_thermal} and \ref{fig:sensitivity_coking}). Domestic production, and supply from different import sources are affected differently by future changes in consumption, however. 

For thermal coal, our model suggests that total imports will see lower growth, or bigger reductions, than total consumption (Fig. \ref{fig:sensitivity_thermal}). Overland imports from Russia are expected to remain relatively stable, whilst imports from Mongolia would grow strongly in percentage terms unless China's thermal coal consumption falls by more than about 1 per cent per year. This is due to the expansion of rail connections into Mongolia, and expansion of Mongolian mines. These overland imports make up a small volume of China's thermal coal imports, however. Seaborne imports from Indonesia and Australia, China's biggest sources of thermal coal imports, are expected to see growth rates of several percentage points below consumption growth rates in any scenario. In total quantity terms, increases or decreases in domestic production cover the bulk of changes in domestic consumption (Supplementary Fig. \ref{fig:scenariosMtinclChina}), whilst reductions in seaborne imports constitute the bulk of reductions in imports in Mt terms (Fig. \ref{fig:sensitivity_thermal}).

Beyond a general relative reduction in total consumption due to a continuing improvement in power plant efficiency, demand for seaborne imports suffers from the combination of an expansion of transport infrastructure that lower the cost of delivery of domestic coal, a relative shift of thermal coal consumption away from coastal and towards inland provinces, and expected development of Chinese mining capacity. These seaborne imports are set to fall unless Chinese consumption would grow at 2 or 3\% pa or more; a level of growth that is beyond the range of most future projections \cite{Auger2021TheAssets}. 

\pagebreak

For coking coal, the increasing availability of steel scrap reduces total demand for primary steel (made with iron ore and coking coal). Even if crude steel production remains at 2019 levels for the foreseeable future, which would be a very bullish scenario, demand for coking coal would fall by about 10 Mt by 2025, and about 45 Mt by 2030, because of the increasing supply of steel from recycled scrap. 

Total imports of coking coal are set to grow, due entirely to growing imports from Mongolia (Fig. \ref{fig:sensitivity_coking}). This is in turn almost entirely due to expanded production at the New Tavan Tolgoi mine, a low-cost producer of high quality hard coking coal. Two rail lines, one 30 Mt rail line connecting the mine to the border crossing at Ganqimaodu in China, and a 10 Mt rail line connecting it to the Russia-China railway at Zuunbayan in Mongolia, will start operations in mid 2022 having been delayed by COVID. Our model suggest these rail lines will be utilized at full capacity, with the much lowered costs of rail vs. truck transport from this mine to the Chinese market making it a very competitive supplier of coking coal.

Overseas imports, predominantly those from Australia, are again expected to see sizeable reductions over time unless consumption in China were to grow at implausibly rapid rates. The rates of change of imports are several percentage points below consumption growth in any scenario. 

\bigskip
Lastly, in our results for 2030 (panels c\&e in Fig.\ref{fig:sensitivity_thermal}\&\ref{fig:sensitivity_coking}), the reductions in imports follow a very similar pattern, and are only slightly more pronounced. This suggests that, while a number of overseas coking coal mines are within the range of the marginal suppliers, there is also a group of overseas mines that will remain competitive in China's seaborne market even when consumption keeps falling substantially. 

\section*{Scenarios for Chinese market and infrastructure development and effect on imports}
\begin{figure}[ht] % hbtp for here, top, bottom, next available page. H for really just here
    \centering
        \includegraphics[width=1\textwidth]{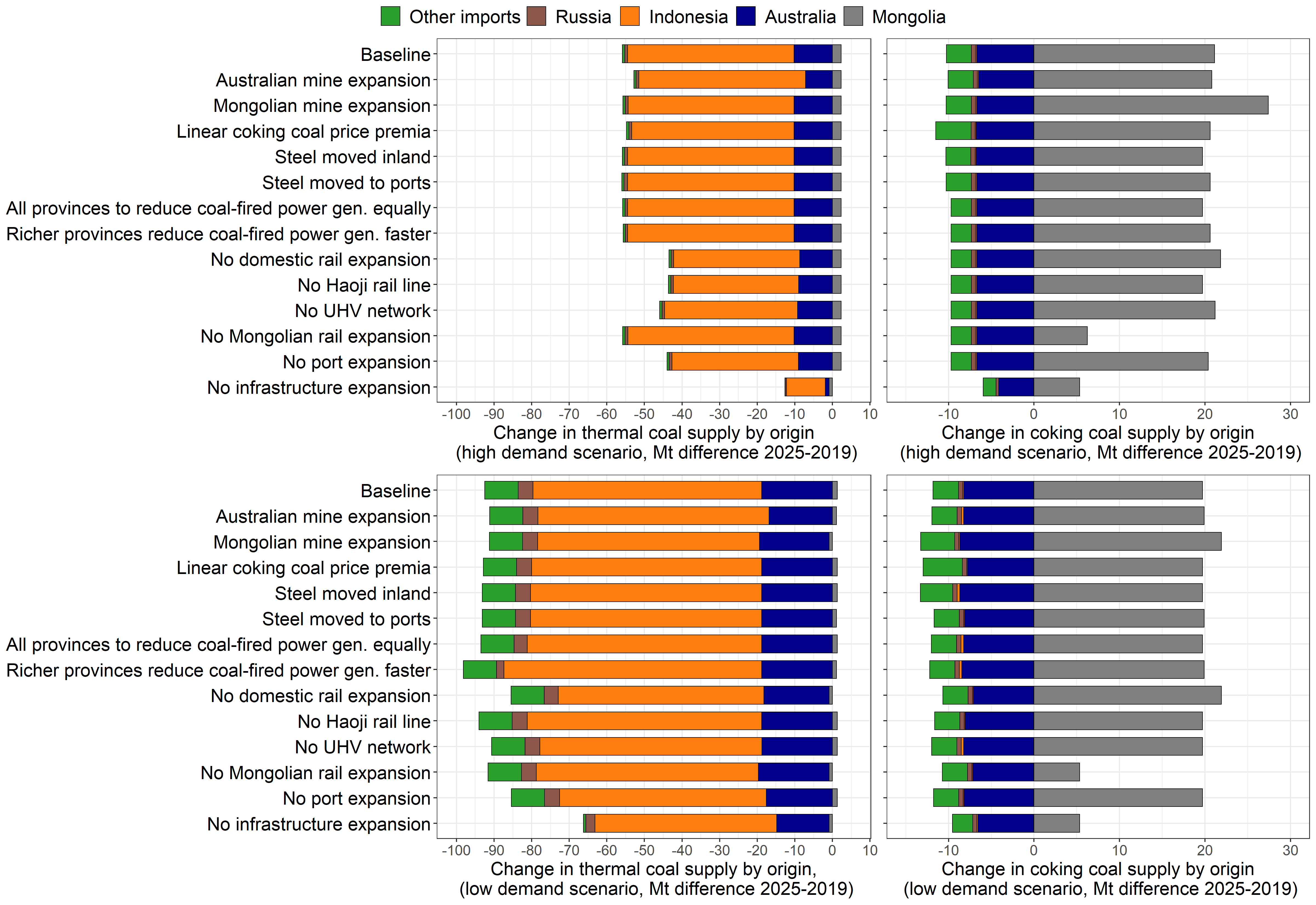} 
    \caption{Scenarios for Chinese market and infrastructure development and effect on imports. 
    \textmd{For scenario definition see the Methods section. These results are presented as total Mt imports in Supplementary Fig. \ref{fig:scenariosMt2025}}}
    \label{fig:scenarios}
\end{figure}

We further investigate how a number of changes to market or infrastructure development would affect resulting imports of coal. We compare levels of imports for each of these scenarios, at both a high and low levels of future consumption growth. These levels of future growth correspond with the IEA's `Stated Policies Scenario' and `Sustainable Development Scenario', respectively, which represent the upper and lower ends of a suite of future energy demand projections, as noted by Auger et al \cite{Auger2021TheAssets}.

Our baseline scenario is described in the previous section, with results presented in Fig. \ref{fig:sensitivity_thermal} and \ref{fig:sensitivity_coking}. Two alternative scenarios include possibilities for future expansions of Australian and Mongolian mines, one scenario uses different assumptions on price premia for different qualities of hard coking coal, and four scenarios allocate future Chinese power and steel consumption differently over different provinces. Lastly, we include six infrastructure scenarios, in which we remove specific elements of China's recent and ongoing infrastructure investments. That is: these infrastructure scenarios do not make assumptions about future infrastructure beyond projects that are currently well under construction, but are rather counterfactual scenarios, in which we ask what future imports would be in the absence of these infrastructure investments. Detailed scenario descriptions are included in the Methods section.

Results shows remarkable similarity of imports across most of these different scenarios (Fig. \ref{fig:scenarios}). Shifts in import levels are driven by the level of consumption to a far greater extent than by the differences between most of our scenarios, at either high or low levels of consumption.

The expansion of production capacity of Australian mines is expected to mitigate reductions in Chinese imports only marginally. Expansion of Mongolian mine capacity would increase Chinese imports of coking coal, by between 2 and 6 Mt/yr, depending on consumption levels.

Whether only coking coals of very high quality are subject to a strong price premium, as we assume in our baseline, or whether prices are more linearly related to coking coal quality, would only very marginally affect imports of Australian and other coking coal. Shifting steel production to major ports benefits overseas importers, whilst moving it toward inland provinces benefits domestic coking coal producers, but again the differences are marginal, and even barely noticeable when demand for steel remains high. Whether reductions in coal-fired power generation will be most rapid in the richer, coastal provinces, or relatively more rapid in inland provinces, may cause a difference of about 10 Mt in Indonesian thermal coal imports in the high demand scenario; in the low demand scenario market losses are similar regardless of where in China the consumption is reduced.

The largest differences to coal imports are those caused by China's infrastructure investments. The expansion of domestic rail lines, the HaoJi rial line specifically, the UHV networks, or port expansions, each had their effect on thermal coal imports. The combined effect of these infrastructure investments is the most striking. In the case of low demand development, these investments will prevent about 25 Mt of additional imports. In the high demand scenarios, overseas imports would still be very similar to 2019 levels in 2025, had the Chinese government not invested in coal transport infrastructure. 

For coking coal imports, these infrastructure investments have much more limited effects, apart from the expanded rail capacity into Mongolia. Those rail lines enable an increase of approximately 15 Mt of imports. Our model also suggests that even if the rail lines to the New Tavan Tolgoi mine had not been built, coking coal imports would increase by about 6 Mt from 2019 levels by 2025. We should note, however, that our model does not impose technical limits on the volume of coal that can be trucked across the border. For import of Australian coking coal, there is a difference of about 2 Mt between the most limited and most expansive infrastructure investments scenarios.

Results for 2030 are similar though more pronounced, in particular for reductions in thermal coal imports (Supplementary Fig. \ref{fig:scenarios2030}\&\ref{fig:scenariosMt2030}).

\section*{Conclusion and discussion}
China has grown to be the world's largest consumer and importer of thermal and coking coal. Its recent and future plans for energy security and decarbonization have, and will continue to, affect the consumption of seaborne imports of coal in particular.

Using a new model that has far greater geo-spatial and technical detail than comparable models, harnessing a wide range of data sources China's, we have assessed scenarios for China's future coal imports. Results show that Chinese imports of overseas coal are likely to fall substantially, in particular imports of Indonesian thermal coal, and Australian thermal and coking coal, even under moderate reduction in Chinese coal consumption. Accelerated decarbonisation, as well as recent infrastructure investments, and expansion of mine capacity in Mongolia, will amplify this trend.

Seaborne imports of thermal coal are expected to fall between 52 and 96 Mt by 2025 and between 56 and 124 Mt by 2030, depending on how rapidly, and where in China, decarbonization occurs. For coking coal, seaborne imports are expected to fall between 9 and 13 Mt by 2025 and between 12 and 17 Mt by 2030, whilst overland imports of coking coal from neighbouring Mongolia are expected to surge by circa 20 Mt by 2025, and remain at roughly that level through 2030. The key factor is China's recently completed or committed domestic coal transport infrastructure. If China had not substantially invested in its coal transport infrastructure, imports in 2025 and 2030 would have remained at roughly similar levels to 2019 in high future demand scenarios, and would have fallen substantially less in low future demand scenarios.

This has clear implications for exporters in the Asian seaborne coal market, primarily thermal coal mines in Indonesia and thermal and coking coal mines in Australia. The expected drop in Chinese demand for seaborne imports in even the highest of future growth projections, should be considered in determining the future value of existing and planned mines, and in the predicted future revenue from coal export royalties and taxes flowing into government budgets for key exporting countries in the region.

The model presented here solves for optimal, i.e., lowest total cost of production and transport, as is usual for these types of model. It does not consider factors such as politically imposed limits on imports from all or from certain countries, such as a recent halt on coal imports from Australia. Including such considerations would require expanding the model to include representation of other key customers in the Asian coal market, including Japan, South Korea, India and Vietnam. 

Our baseline scenario does not include any additional infrastructure development beyond projects that are well under construction. Whilst it may be possible that further investment in rail lines to low-cost coal producing regions in China's West would push out more overseas suppliers, it is also likely that China has reached 'peak coal infrastructure', and that government planners won't see great benefits from developing further big rail projects in a time of declining overall consumption. 

Lastly, our model finds cost optimal supply for annual levels of demand, with restrictions in annual transport capacities, again as is usual for these types of model. This does not very well account for temporary and local demand peaks, caused for example by the seasonality of electricity demand and volatility of China's hydropower output, or for temporary supply disruptions such as caused by mine flooding or COVID-19 related production or transport restrictions. Whilst Chinese economic planners are demanding increased inventory stockpiles at power plants and key coal logistics hubs, in response to recent supply shortages leading to rationing of power supply in a number of Southern provinces, such temporary spikes in supply shortages may mitigate the effects of some of the demand reductions on China's coal imports predicted here.

\pagebreak
\section*{Methods}

\subsection*{Model description}
Our model is a linear optimization problem for coal supply to the Chinese coal market, with an objective function that minimizes the sum of production (mining) and transportation costs. The model includes separate demand levels for electricity from coal-fired power, thermal coal use in other industries, and coking coal use in steel making. This demand is inelastic and must be met. It includes technical constraints on the capacity of coal-fired power plants and steel plants, railway lines, and ports, and considers the conversion efficiency of coal-fired power plants. Model inputs on coal mining capacity, infrastructure development, and future demand taken from projections exogenous to the model (more below). The model solves for individual years; for the results for 2025 and 2030 we do account for mine depletion by cross-checking that all mines producing in 2025 or 2030 had six or eleven years worth of modelled production levels in them in the year 2019. The mathematical formulation of the model is provided at the end of this section.

The system of supply, transport and consumption in the model is represented as a node \& link model. The key contribution of our model is the strongly improved granularity of the coal transport network that we build; individual components of this network and how they are strung together are described below (section `network construction and data collection'). 

All links in the model are transport links for physical amounts of coal, with constraints defined in Mt, whilst demand for coal-fired power and other thermal coal use is defined in PJ. Different thermal coals are grouped in calorific value bins, with steps of 250 kcal/kg. 

Demand for coal-fired power is placed in provincial-level electricity demand nodes, and can be satisfied by power plant units within the province, or by power transmitted over the inter-provincial UHV network.  

Demand for thermal coal for other purposes is determined as PJ of primary energy, and is placed at city-center nodes.

Demand for steel is placed in provincial level steel demand nodes and is defined in Mt of crude, primary steel. We consider a mix of hard coking coal (HCC), semi-soft coal (SCC), and pulverized coal for injection (PCI), needed to produce a tonne of steel, with the same mix used by all steel plants. Only steel plant nodes are connected with these steel demand nodes. 

Our model is written in Python, with linear problem formulation done using the package PuLP (\url{pypi.org/project/PuLP}). This allows for replication or development with fully free and open source software. The size of the network in our model does mean it requires fairly substantial computing resources. We use a commercial cloud computing service to process input data on demand levels and model constraints to create a problem definition (.lp) file. For our model, this process requires about 10 hours on an instance with 72 virtual CPU and 144 Gb of RAM. We then download the lp file and solve locally using the CPLEX solver via IBM's CPLEX Interactive. Either step can also be done locally in Python, albeit considerably slower; scripts for both alternative routes are included in our public data and code repository.

\subsection*{Mathematical formulation of the model}
\label{section:formulation}
The sets, parameters, and variables included in the model are listed in \cref{tab:modelsets,tab:modelparameters,tab:modelvariables}.

The objective function \cref{eqn:obj} minimizes the total production and transport of coal, plus the cost of transmission of electricity via the inter-provincial UHV network. The cost of transport over each link is made up of a fixed handling cost, plus a variable distance based component. The variable transport cost varies by transport mode, whilst the handling charges vary with switches between different transport modes (e.g., from mine to truck or from mine to rail). Transmission costs over the UHV network are pre-calculated, and are based on the distance of the transmission link. We apply a conversion efficiency factor to each UHV link to represent line losses, which are based on line length and type of UHV connection (DC or AC). 

The model includes the following constraints. Firstly, a supply constraint \cref{eqn:supplycons} that limits the supply of any type of coal by any node to below or equal to the production capacity for that type of coal.

The mass balance \cref{eqn:massbalance} states that the supply plus transport flows into a node, must be at least equal to the transport flow out of that node, for each type of coal. The energy balance \cref{eqn:nrgbalance} states that the total energy content of the supply plus transport flows into a node, must be at least equal to the energy demand for electricity generation, plus the energy demand for thermal coal for other purposes, plus the total energy content of the transport flows out of that node.

The transport capacity constraint \cref{eqn:transpcapa} limits the total volume of coal transported over each link to its transport capacity; \cref{eqn:portcapa} constrains volumes flowing out of port nodes to port handling capacity. 

We represent conversion losses in power generation by multiplying the the energy content of the different types of coal arriving in power plant unit nodes with the power plant unit's conversion efficiency. The electrical generation capacity of these power plant units is represented in \cref{eqn:eleccapa} by imposing a transmission capacity on links between power plant units and provincial-level electricity demand nodes, which is equal to the total energy content of the coal, multiplied by the conversion efficiency of the power plant unit; the same logic is applied to links in the UHV network. 

We represent the process of steel making by requiring the links between steel plants and steel demand centers to transport a volume of coking coal required for the production of the demanded volume of steel. \cref{eqn:stptcapa} limits the flow of coking coal over these links to the amount of coking required for the steel plant to run at full production capacity. The hard coking coal balance \cref{eqn:hccbalance} states that all steel demand must be satisfied, and simultaneously that this should be done with the required 581 kg of HCC for every ton of steel produced. \cref{eqn:sccbalance,eqn:pcibalance} dictate that the required mix of coking coals, with 581 kg of HCC, 176 kg of SCC, and 179 kg PCI, is used.

\begin{table}[H] % hbtp for here, top, bottom, next available page. H for really just here
\caption{Model sets}
\label{tab:modelsets}
\begin{tabularx}{\textwidth}{llX}
\toprule
Set   &   Description    &   Notation \\ \midrule
(\textit{i,j})   &   Set of all nodes    &   \textit{$\in$ I}\\
\textit{p}   &   Subset of nodes that are ports    &   $\in$ \textit{P} $\subset$ \textit{I}\\
\textit{s}   &   Subset of nodes that are steel plants    &   $\in$ \textit{S} $\subset$ \textit{I}\\
(\textit{k,k'})   &  Index of node types    &   $\in$ \textit{K} = [Mine, basin, railway stop, city center, port, navigation waypoint (river), navigation waypoint (ocean), power plant, power plant unit, steel plant, provincial power demand center, provincial steel demand center] \\
\textit{t}   &   Years    &   $\in$ \textit{T} = [2015,$\cdots$,2030] \\
\textit{c}   &   Coal types    &   $\in$ \textit{C} = [Thermal coal, HCC, SCC, PCI] \\
\textit{v}   &   Bins with average calorific value of coal types    &   $\in$ \textit{V} = [0, 3000, 3250,$\cdots$, 7000]\\
\textit{m}   &   Transportation modes    &   $\in$ \textit{M} = [Rail, truck, river barge, ocean-going ship] \\
\midrule
\end{tabularx}
\end{table}

\begin{table}[H] % hbtp for here, top, bottom, next available page. H for really just here
\caption{Model parameters}
\label{tab:modelparameters}
\begin{tabularx}{\textwidth}{llX}
\toprule
Parameter   &   Unit    &   Description \\ \midrule
elec\textunderscore demand$_{\textit{i,t}}$   &   PJ    & Demand for coal-fired power generation in node \textit{i} and year \textit{t} (as energy content of the electricity generated) \\
other\textunderscore demand$_{\textit{i,t}}$   &   PJ    & Demand for thermal coal for other uses in node \textit{i} and year \textit{t} \\
steel\textunderscore demand$_{\textit{i,t}}$   &   Mt    & Demand for steel in node \textit{i} and year \textit{t} \\
CV$_{\textit{v}}$     &   kcal/kg    & Calorific value associated with thermal coal in bin \textit{v}\\
kcal\textunderscore to\textunderscore PJ &   --    & Conversion factor for energy content, from kcal/kg to PJ/Mt \\
HCC$_{\textit{c}}$     &   --    & Dummy variable indicating coal type is hard coking coal \\
SCC$_{\textit{c}}$      &   --    & Dummy variable indicating coal type is semi-soft coking coal \\
PCI$_{\textit{c}}$      &   --    & Dummy variable indicating coal type is pulverized coal for injection \\
prod\textunderscore cost$_{\textit{i,c,t}}$      &  \$/t    &  Production cost of coal type \textit{c} at node \textit{i} and year \textit{t}\\
prod\textunderscore capa$_{\textit{i,c,t}}$      &  Mt    &  Production capacity of coal type \textit{c} at node \textit{i} and year \textit{t}\\
transp\textunderscore capa$_{\textit{i,j,t}}$      &  Mt    &  Transport capacity between nodes \textit{i} and \textit{j}  in year \textit{t}\\
port\textunderscore capa$_{\textit{i,t}}$      &  Mt    &  Maximum handling capacity of port in year \textit{t}\\
elec\textunderscore capa$_{\textit{i,j,t}}$      &  PJ    &  Electrical transmission capacity between between nodes \textit{i} and \textit{j} in year \textit{t}\\
stpt\textunderscore capa$_{\textit{p,t}}$      &  Mt    &  Production capacity of steel plant \textit{p} in year \textit{t}\\
conv\textunderscore eff$_{\textit{i,j}}$      &  --    &  Efficiency of electrical conversion or transmission between nodes \textit{i} and \textit{j}\\
distance$_{\textit{i,j}}$      &  km    &  Distance between nodes \textit{i} and \textit{j} \\
transp\textunderscore cost$_{\textit{m,t}}$      &  \$/t$\cdot$km    &  Variable cost for a transport mode \textit{m} in year \textit{t}\\
handling\textunderscore cost$_{\textit{k,k',t}}$      &  \$/t    &  Fixed handling charge for transport between node of type \textit{k} and type \textit{k'} in year \textit{t}\\
transm\textunderscore cost$_{\textit{i,j,t}}$      &  \$/PJ    &  Variable cost for a transmission of electrical energy between nodes \textit{i} and \textit{j} in year \textit{t}\\
\midrule
\end{tabularx}
\end{table}

\begin{table}[H] % hbtp for here, top, bottom, next available page. H for really just here
\caption{Model variables}
\label{tab:modelvariables}
\begin{tabularx}{\textwidth}{llX}
\toprule
Variable   &   Unit    &   Description \\ \midrule
supply\textunderscore Mt$_{\textit{i,c,t}}$      &  Mt    &  Production of coal type \textit{c} at node \textit{i} and year \textit{t}\\
transp\textunderscore Mt$_{\textit{i,j,c,t}}$      &  Mt    &  Transport of coal type \textit{c} between nodes \textit{i} and \textit{j} in year \textit{t}\\
\midrule
\end{tabularx}
\end{table}

\bigskip
The objective function:
\begin{equation}
\label{eqn:obj}
\begin{aligned}
\textit{Minimize}\sum_{i,c,t} supply\_Mt_{i,c,t} \times prod\_cost_{i,c,t} \times 1 \cdot e^{6} + \\
\sum_{i,j,c,t}transp\_Mt_{i,j,m,t} \times (handling\_cost_{k,k',t} + transp\_cost_{m,t} \times distance_{i,j}) \times 1 \cdot e^{6} + \\
\sum_{i,j,v,t}transp\_Mt_{i,j,v,t} \times CV_{v} \times 1 \cdot e^{9} \times kcal\_to\_PJ \times conv\_eff_{i,j} \times transm\_cost_{i,j,t}
\end{aligned}
\end{equation}

\bigskip
The model constraints:

\bigskip
Coal mine production capacity:
\begin{equation}
\label{eqn:supplycons}
0 \geq Supply\_Mt_{i,c,t} \leq prod\_capa_{i,c,t}
\end{equation}

Mass balance:
\begin{equation}
\label{eqn:massbalance}
Supply\_Mt_{i,c,t} + \sum_{j,i,c,t}transp\_Mt_{j,i,c,t} \geq \sum_{i,j,c,t}transp\_Mt_{i,j,c,t}
\end{equation}

Energy balance:
\begin{equation}
\label{eqn:nrgbalance}
\begin{aligned}
Supply\_Mt_{i,c,v,t} \times CV_{v} \times 1 \cdot e^{9} \times kcal\_to\_PJ + \\
\sum_{j,i,c,v,t}transp\_Mt_{j,i,c,v,t} \times CV_{v} \times 1 \cdot e^{9} \times kcal\_to\_PJ \geq \\
elec\_demand_{i,t} + other\_demand_{i,t} + \\
\sum_{i,j,c,v,t}transp\_Mt_{i,j,c,v,t} \times CV_{v} \times 1 \cdot e^{9} \times kcal\_to\_PJ
\end{aligned}
\end{equation}

Transport capacities:
\begin{equation}
\label{eqn:transpcapa}
\sum_{i,j,c,t}transp\_Mt_{i,j,c,t} \leq transp\_capa_{i,j,t}
\end{equation}

Port capacities:
\begin{equation}
\label{eqn:portcapa}
\sum_{p,j,c,t}transp\_Mt_{p,j,c,t} \leq port\_capa_{p,t}
\end{equation}

Electric capacities:
\begin{equation}
\label{eqn:eleccapa}
\sum_{i,j,c,v,t}transp\_Mt_{i,j,c,v,t} \times CV_{v} \times 1 \cdot e^{9} \times kcal\_to\_PJ \times conv\_eff_{i,j} \leq elec\_capa_{i,j,t}
\end{equation}

Steel plant capacities:
\begin{equation}
\label{eqn:stptcapa}
\sum_{s,j,c,t}transp\_Mt_{s,j,c,t} \leq stpt\_capa_{s,t} \times 0.966
\end{equation}

Hard coking coal balance:
\begin{equation}
\label{eqn:hccbalance}
\sum_{p,j,c,t}transp\_Mt_{p,j,c,t} \times HCC_{c} \geq steel\_demand_{j,t} \times 0.581
\end{equation}

Semi-soft coking coal balance:
\begin{equation}
\label{eqn:sccbalance}
transp\_Mt_{p,j,c,t} \times HCC_{c} \div 0.581 = transp\_Mt_{p,j,c,t} \times SCC_{c} \div 0.176
\end{equation}

PCI balance:
\begin{equation}
\label{eqn:pcibalance}
transp\_Mt_{p,j,c,t} \times HCC_{c} \div 0.581 = transp\_Mt_{p,j,c,t} \times PCI_{c} \div 0.179
\end{equation}

\subsection*{Model network construction and data collection}
\label{section:datacol}
The defining quality of our model is the granularity of the network that we use to represent China's system of production, transport, and consumption of coal; see the overview in Table \ref{tab:nodes-links} and Fig. \ref{fig:maps}. Larger scale maps are provided in Supplementary Fig. \ref{fig:rail-network}-\ref{fig:power-steel-plants-map}. We rely on a number of open and proprietary data sources for the different components of this network.

For our railway network, we rely on China Railway Map (\url{cnrail.geogv.org}), a project that in turn builds on data from OpenStreetMap (\url{openstreetmap.org}) and Chinese railway schedules. This website provides names of each line, its type (freight, passenger, or combined), names and approximate location (county name) of railway stops, and distance between stops. We scraped the Chinese names and locations of all stops, and matched with data in OpenStreetMap to find latitude and longitude of each. Where no exact match was found, we queried Google's geocoding API for the location of the railway stop, with Chinese names and county name as the search string, using the ggmap package (\url{github.com/dkahle/ggmap}) in R. We plotted each of the 145 railway lines individually and visually inspected approximate location and route of each line versus the China Railway Map. Where the geocoding had resulted in imprecise locations, we manually fixed latitude and longitude based on visual comparison with the China Railway Map. For transport capacities of each line and year we relied on a mix of official government reports, public investor briefing documents, and data from Wood Mackenzie. Where rail lines were connected with ports that were found to handle coal, we adjusted the capacity of the rail line to the handling capacity of the port, more details below. For a small number of lines we added extensions or capacity upgrades based on information in government documents

For our road network, we started with a list of China's 685 cities (including provincial, prefecture, and county-level cities). We used Google's geocoding API and ggmap to determine latitude and longitude of these cities. We then calculated geodesic ('as the crow flies') distances between each pair of cities, and preserved the nearest 12 cities for each city. We then queried actual driving distances between this collection of city pairs, using Google's distance matrix API via ggmap, and retained the nearest eight connections for each city. Google's distance matrix could not provide driving distances between locations in Hong Kong, Macau and mainland China, so we used geodesic distances multiplied by 1.7, the average for other links in our dataset.

For our navigation network, we picked navigation waypoints roughly every 100 km along China's coastline and calculated geodesic distances between them. Ports and coastal power or steel plants with ports were connected with the nearest such waypoint, again with geodesic distance between them calculated. For ports along rivers and their connection with navigation waypoints, we calculate geodesic distance and multiply with 1.8, the approximate 'sinuosity' (the factor of navigational to geodesic distance) of the Yangtze river. This sinuosity was based on navigation distances as reported by Kpler between the Yangtze river mouth and the farthest upstream port. We use the same sinuosity for other rivers, as there are only a very small number of power or steel plants rivers other than the Yangtze. Based on Kpler offloading data, we consider ocean-going vessels to be able to sail up the Yangtze river no further than the bridge at Jiangyin (at coordinates 31.944, 120.274). Coal from such vessels can only be transported beyond this point after being loaded onto a river barge at one of the coal ports in the Yangtze river mouth.

We include a UHV transmission network, using an overview of lines that are operational or under construction as created by the Lantau group \cite{TheLantauGroup2020CuttingEvolves}. We use the line length and transmission capacity as provided in this report. Because we are interested in coal-fired power generation, we set this transmission capacity to zero if the Lantau report states that the dominant generation source at the origin is renewables or nuclear, 50\% of capacity if the report mentions coal and other sources, and 100\% if the cable originates from coal-rich areas of Shaanxi, Shanxi, or Inner Mongolia, or if no explicit generation source at the origin is mentioned.

For ports, we use the Kpler (\url{kpler.com}) database for dry bulk goods, which tracks individual vessels and provides estimates of quantities and types of goods loaded and unloaded, based on port authority websites, amongst others. This dataset includes 274 ports and provides data back to the year 2017. We compared Kpler maps with Google maps to get location data for these ports. We used a three-month rolling average to determine port capacity and expansions over time, and use the earliest estimate for the years 2015 and 2016. This data was compared with publicly available data on official handling capacities for major coal loading ports in the north of China, but we found no reason to make any corrections. 

For power plants and steel plants, we rely on the global power plant tracker and the global steel plant tracker developed by Global Energy Monitor (\url{globalenergymonitor.org}). The developers of this initiative kindly made the full dataset for Chinese power and steel plants available to us. These datasets include data on 5,537 units at 1,693 power plants, and 305 steel plants, including plants or units that are retired, operational, under construction, or planned. For each, this dataset provides start of operations, annual production capacity, precise location, and an estimate of conversion efficiency for power plant units. We used the map view on the tracker websites to verify whether power and steel plants along the ocean and rivers had their own offloading port, whether they were located in bigger ports, and whether they had a rail connection. This added another 190 port nodes for power plants, and 24 port nodes for steel plants. 

For mining locations, we use the coal supply dataset for China and Mongolia from Wood Mackenzie. This includes 174 Chinese mine clusters and 25 individual Mongolian mines. The data does not include a precise location for the Chinese mine clusters. Instead we use provincial level maps provided by Wood Mackenzie to locate coal basins within provinces. We then include direct connections (calculating only handling charges but zero distance based charges) to all rail lines that cross these basins, and approximate trucking distances between basins and cities, using a single value for each province, based on a rough estimate of average distance from basin to the edges of the province. 

We then add links between the different types of nodes. For power plant and steel plants, we calculate geodesic distance to all railway stops, and preserve all stops within 25 km. We then query driving distances to each of these stops with the ggmap package in R, and preserve only the nearest stop on each distinct rail line. Where the geodesic distance was less than 5km, we presume a direct rail link between the stop and plant. For stops within 15km driving distance, we include a direct trucking link, and where driving distances are longer we discard the link and presume trucking to occur via a city center node. Approximately 30\% of power plants was within 5km of a railway stop, and about 73\% within 15km. We then visually inspected satellite images of all power plants that were further than 50km from a railway stop to identify mine-mouth plants; this resulted in 43 such links. We created trucking links between city centers and each railway stop within 15 km; for these links we set driving distance to zero but do include handling costs. We also create trucking links between power and steel plants and the three nearest city centers for each. We visually inspected satellite images of each port to identify what railways they connected with, and created direct links for these. We include trucking links between ports and the nearest three city centers, and between ports and power or steel plants within 50 km driving distance. Foreign mines are linked to all navigation waypoints, with a single point of departure for each country to approximate distances. We use the biggest coal port in each country for that purpose, i.e., the port of Newcastle in Australia, the port of Tarahan in Indonesia, and the port of Vladivostok in Russia. For all mines in all other countries we create similar links and assume one same navigation distance of 12,000 km. We create overland links from Russia from Zabaikalsk to Manzhouli, and from Grodekovo to Suifenhe, the two points at which there are border crossing rail lines. We also create trucking links at these two points. For links between Mongolia and China, we include a representation of Mongolia's coal rail network, the connections with individual mines, and trucking links with distances as estimated by Wood Mackenzie.

Lastly, we create a number of functional links between power plant nodes and power plant unit nodes, between power plant units and provincial-level electricity demand nodes, and between steel plants and provincial-level steel demand nodes. These provincial-level power and steel demand nodes are fictive locations where we place demand, which may then be fulfilled with production in any power or steel plant within that province. The UHV network is represented as links between these provincial-level power demand nodes.

All links in the network are dual direction, apart from all links that go towards power or steel plants, and from power or steel plants into provincial-level demand centers.

\subsection*{Coal mine capacities and coal mining costs}
We use Wood Mackenzies' coal supply data for China and Mongolia, which provides prices, production costs and production levels for 174 Chinese mine clusters and 25 individual Mongolian mines. We use the 'base view' scenario for development of Chinese and Mongolian coal mining capacity provided with this data. This includes investment levels over the period 2020-2025 that is similar to the period 2015-2019, and falls to about half as much over the period 2026-2030 \cite{WoodMackenzie2020China2020}. Resulting development of production capacity is provided in Supplementary Fig. \ref{fig:mine_capa}. An exception is made for the production of Fig. \ref{fig:sensitivity_thermal}, specifically the scenarios for the year 2030, with 4 and 5\% annual growth relative to 2019. The increase in coal consumption in those scenarios exceed total technical capacity in Wood Mackenzie's base view for 2030. For these two demand scenarios only, we allow mines to produce the same level as their 10 year running maximum, provided they are not depleted yet, and at the same costs as when they were running at their running maximum capacity. We do not consider alternative scenarios with smaller or greater amounts of investment in mine capacity. 

For foreign suppliers, we use global supply curves for thermal and coking coals with mine level production levels and costs, again from Wood Mackenzie. Because our model does not include demand for coal outside of China, we need to correct this curve to reflect the fact that Chinese customers will not be able to buy all of the cheapest coals. We consider the key suppliers to China, and determine the average share of exports to China over 2017-2019, separately for thermal and coking coals. We multiply the width of each supply step (i.e., the production capacity in Mt of each mine) in these key supplier countries with this correction factor; for a visualization and correction factors applied, see Supplementary Fig. \ref{fig:glob_supply_curves}.

We use the global supply curve for 2019 as a proxy for the years 2025 and 2030 for the production costs and capacity of foreign mines. Production costs for most mines in the China and Mongolia data set are expected to be virtually stable in real terms, and using the same trend globally therefore does not seem unrealistic. For investment in mining capacity outside of China, this effectively means that we consider mine depletions and retirements to be balanced out by newly opened capacity. Whilst this is a somewhat simplistic assumption, this choice can not be expected to be driving our results; new production capacity outside of China will almost certainly have similar or higher production costs than is currently the case; otherwise such coal resources would already be exploited. Note also that we do consider two scenarios in which we include additional expansion of either Australian or Mongolian mines.

\subsection*{Coal quality and price premia}
We include considerations for coal quality, and how these characteristics affect the competitiveness of different coal types. 

For thermal coal, we use data from the Wood Mackenzie coal supply dataset on prices and qualities including calorific value (CV), levels of ash, volatile matter, sulphur content etc. We use regression analysis to determine the relationship between these coal prices and qualities. We find no significant effect on prices of any of these coal qualities, apart from the calorific value. This is true even after standardizing to a specific CV value, i.e., a coal type with double the CV value have a price that is more than double. Our regression analysis showed that this price premium was circa \$8 per 1000 kcal/kg; see Supplementary Fig. \ref{fig:premiathermal}. 

For coking coal, we find that the only quality that significantly drives prices is the `Coking Strength after Reaction', or CSR number. We regard global prices paid for hard coking coals and their CSR, and find a non-linear relationship with strong price premia paid for coking coals with very high CSR in particular; see Supplementary Fig. \ref{fig:premiacoking}.

We reflect the market value of these different coal types by subtracting these price premia from mine gate costs. That is, our model optimizes for minimum cost minus price premia.

\subsection*{Transport costs}
In China, transport costs via rail are set by the state, and have been at the same level since 2015, with a fixed handling charge of 16.3 CNY/t, and a distance based charge of 0.131 CNY/t$\cdot$km \cite{NDRC2017Notice2163}. A number of rail lines are operated by integrated coal mining and power generation companies \cite{Rioux2016}; we assume the same costs for those. The dedicated coal rail lines Mengji, Wari, and Haoji/Menghua have a distance based charge of 0.184 CNY/t$\cdot$km and zero handling costs; the Shuohuang line has a distance based charge of 0.12 CNY/t$\cdot$km and handling costs of 16.3 CNY/t. We use distance based charges of 0.25 CNY/t$\cdot$km for trucks, 0.08 CNY/t$\cdot$km for river barges, and 0.02 CNY/t$\cdot$km for ocean-going ships \cite{WoodMackenzie2020Transport2020}. We assume 25 CNY/t port handling costs, and use the national railway handling charge of 16.3 CNY/t as a guide for handling charges between other types of transport. For the UHV line network, we use a generic transmission cost of 35 CNY/MWh$\cdot$1,000 km and line losses of 2.8\%/1,000 km for UHV-DC lines and 3.6\%/1,000 km for UHV-AC lines \cite{CCIDConsulting2020WhiteConstruction}. 

\subsection*{Locating coal consumption}
\label{section:consumption}
For our future scenarios, and for historical calibration, we distribute national levels of coal consumption to the provincial level, and then distribute provincial level coal consumption over individual power and steel plants, or city-level nodes. To determine the location of this consumption of coal, we start with historical data from SXcoal (\url{SXcoal.com}), which relies on official statistics from government and customs data, and splits consumption at the national level into power generation, steel making, construction materials, chemicals, heating, and other uses.

For power generation, SXcoal also provides provincial-level numbers for `thermal' power production, a Chinese statistical category that was 90.2\% coal-fired power generation at the national level in 2019 \cite{ChinaElectricityCouncil2020Basic2019}. We subtract gas-fired power generation from provincial levels of thermal power generation, by using 2019 data on gas-fired generation capacity by province \cite{Zhitan2020ThoughtsPlan} as a proxy for electricity generation by province. We assume the same distribution over provinces, as a percentage of national gas-fired power generation, for the years 2015 to 2019. We use the same procedure to subtract provincial level biomass-fired power generation \cite{BiomassObserver2020TheReleased}. Other generation sources make up about 2.5\% of the `thermal' category; we subtract this from provincial-level thermal power generation using that same percentage for all provinces. This leaves us with estimates of coal-fired power generation per province over the period 2015-2019. In our network, we place total demand for electricity from coal-fired power generation in provincial-level demand nodes, but the coal is consumed and converted to electricity in the power plant nodes. 

For steel making, we derive a mix of coking coal consumed per tonne of primary steel produced with data from Wood Mackenzie\cite{WoodMackenzie2021China2021}. This data provides Chinese and Mongolian supply data for three different types of coking coal (HCC, SCC, and PCI). We add imports of coking coal (data from SXcoal), with estimates of the type of coking coal of these imports. We use crude steel production data, and subtract scrap consumption data, to calculate primary steel production\cite{WorldSteelAssociation2021World2020}. This gives very stable estimates of coking coal consumption per tonne of primary steel, at 581 kg HCC, 176 kg SCC, and 179 kg PCI (average over 2017-2019); see Supplementary Fig. \ref{fig:mix_coking_coal}. We use the same mix of coking coal consumption per ton of primary steel for future years, and for all steel mills. As with power production, we place total demand for steel in provincial-level demand nodes, but the coal is consumed and used to produce steel in the steel plant nodes.

For construction materials, we use SXcoal data on provincial-level production of cement as a proxy to split the national level consumption of coal over individual provinces. We further disaggregate this consumption over individual cities within each province, using plant-level emissions of cement plants in 2019 from the Global Energy Infrastructure Emissions Database (\url{gidmodel.org/?page_id=27}) as a proxy. Roughly two-thirds of national emissions reported in this dataset is not attributed to individual cement plants. We aggregate the plant-level data to the city-level, and assume the fraction of cement production that is unaccounted for is similarly distributed over cities within each province. We use the same split for the years 2015-2019.

For chemicals production, we use SXcoal data on provincial-level production of ammonia as a proxy to split national level consumption of coal over individual provinces. Ammonia production is responsible for the largest share of emissions from the chemicals industry globally \cite{IEA2020Chemicals.Needed}. We further disaggregate this consumption over individual cities within each province, using city-level GDP, as a share of provincial level GDP, as a proxy for production levels of chemicals in each city.

For heating, we use provincial-level heating degree days \cite{Shen2016ChangesChina} (HDD), multiplied with city-level population numbers, as a proxy for coal consumption, to distribute national-level consumption of coal over city-level nodes. Shen and Liu \cite{Shen2016ChangesChina} do not provide HDD numbers for Beijing, Tianjin, or Shandong, even if they do place these within China's winter heating zone (partially so for Shandong). We use the HDD number for neighbouring Hebei as a proxy for these provinces. We do the same for Shaanxi, using the HDD number for neighbouring Shanxi.

For `other' coal consumption as reported by SXcoal, we use city-level population numbers as a proxy to distribute national-level consumption to city-level nodes.

For the total amount of energy demand used as model input, we substitute data on the consumption of coal in the power generation industry, from SXcoal's national level data on consumption by industry, with SXcoal's data on coal-fired power production, as the latter is split by province. Further, using electricity generated rather coal consumed by the power industry allows our model to consider power plant efficiency in its optimization of coal consumption. The two datasets are not consistent however, with the reported power production exceeding what could logically be generated from the consumed coal. This is likely because the former includes power generated at captive power plants in industries other then the power generation industry. In order to preserve the national level energy balance, we reduce the consumption in `other' uses by the same amount as is added to consumption in the power generation industry. We also substitute coal consumption data in the steel making sector with our own estimates of the coking coal mix described above; this data is perfectly consistent with SXcoal's national level data, however. For details on this energy balance adjustment see Supplementary Tables \ref{tab:balance_orig_Mt}-\ref{tab:balance_adj_PJ}. This adjustment is mostly needed to determine levels and locations of coal consumption in forecasted demand for the years 2025 and 2030.

\subsection*{Forecasts of coal consumption}
For our results on the sensitivity of coal imports from different suppliers to changes in Chinese consumption levels (Fig. \ref{fig:sensitivity_thermal}\&\ref{fig:sensitivity_coking}), we do not make any assumptions on future consumption, but rather set a very wide range of potential growth levels. Note that for results presented for thermal coal in Fig. \ref{fig:sensitivity_thermal}, we set the growth of coking coal consumption to zero, and for results presented for coking coal in Fig. \ref{fig:sensitivity_coking}, we set the growth of thermal coal consumption to zero.

For our scenario comparisons in Fig. \ref{fig:scenarios} we use a high and low demand development levels. We do not develop these demand scenarios ourselves but base these on the IEA’s `Stated Policies Scenario’ and `Sustainable Development Scenario’; exact numbers used are included in Supplementary Tables \ref{tab:IEA_SPS}\&\ref{tab:IEA_SDS}. 

These demand levels are all specified at the national level, however, and require us to make assumptions about how future consumption is distributed over different provinces. 

Different provinces have seen strongly differing growth rates of coal consumption over recent years, largely due to strongly differing levels of development, economic growth, industrial activity, and power consumption per capita, etc., with the biggest differences between coastal and inland provinces. We use the following method to account for these differences in distributing national level consumption over different provinces. For power generation, we calculate the average annual percentage growth rate of coal-fired power generation over the period 2015 to 2019, for each province. Then we add or subtract a certain number of percentage points to each provinces' historical growth rates to extrapolate through to 2025 or 2030, such that sum of extrapolated consumption at provincial level correspond with national consumption levels from our scenario setting (for a visualization see Supplementary Fig. ). For use of thermal coal for construction materials, chemicals, heating, and other uses, which together account for approximately 22\% of all coal consumption, we use a more simple extrapolation method, as we see no clear relation with e.g., economic development levels of different provinces. For these, we take 2019 levels and multiply these with the CAGR for thermal coal consumption for `other uses' as given in the scenarios (Supplementary Tables \ref{tab:IEA_SPS}\&\ref{tab:IEA_SDS}). That is, we use the same growth rate for all provinces and industries, but with the 2019 level for each province and industry as the base level. Note that we test the sensitivity of our results to these choices, by including two scenarios that apply alternative ways to distribute future thermal coal consumption over different provinces.

For steel making, we also use the more simple method of using 2019 provincial-level percentage shares to split future coking coal consumption over different provinces. Again, this is justified as we do not see a clear relation with e.g., economic development levels of different provinces. Rather, steel production is highly concentrated, with  56\% of all steel in 2019 produced in just five provinces, and the same five provinces were responsible for 40\% of the growth in production between 2015 and 2019. Again, to test sensitivity of our results to this choice, we develop a number of alternative scenarios that place steel production growth in different locations.

We note here that both steel and power plant capacity are well more than sufficient to produce national level demand in any of our scenarios, as coal-fired power plant utilization rates (capacity levels) were below 50\% in 2020, whilst steel production capacity exceeds output by about 10\%, and we see limited future growth potential of steel production volumes. We therefore do not consider any investment in such plants. The only exception is a scenario where we move steel production inland. For this scenario, we expand the capacity of existing inland steel plants with a single factor, such that total production capacity for each province exceeds total demand in that province by 5\%. 

\subsection*{Model calibration}
\begin{figure}[h] % hbtp for here, top, bottom, next available page. H for really just here
    \centering
        \includegraphics[width=1\textwidth]{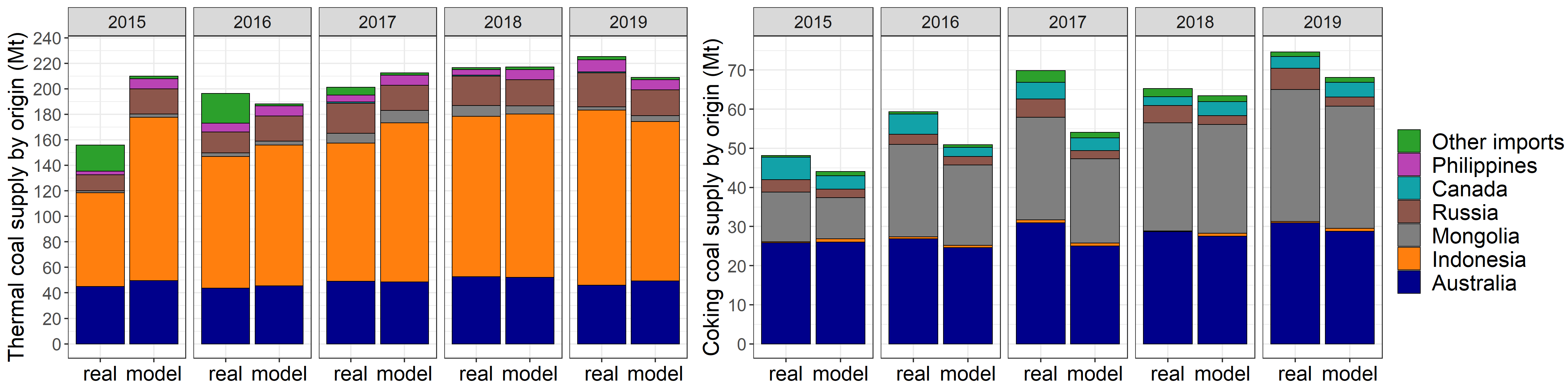} 
    \caption{Model calibration results.
    \textmd{}
    }
    \label{fig:calibration}
\end{figure}

We run our model for the years 2015 through 2019, with historical data on operational rail lines and ports, UHV transmission lines, and power and steel plants. Modelled results closely track actual total reported imports, as well as their split by origin \ref{fig:calibration}, which gives confidence about the validity of predicted outcomes for future years. Crucially, our model does not appear biased against foreign imports, and therefore our future projections of reduced demand for foreign coals can not be attributed to model design. 

Note that in our results in Fig.\ref{fig:sensitivity_thermal}, \ref{fig:sensitivity_coking}, and \ref{fig:scenariosMt2025}, we list 2019 supply levels and changes in supply levels versus 2019, as changes versus modelled results for 2019. Although very close to real levels, modelled outcomes for 2019 are still somewhat lower, and using the latter to calculate changes in supply removes any effect modelling choices may have had on outcomes. 

\subsection*{Scenario definition}
Baseline: our baseline scenario includes production capacity and costs for individual mines, and projections of power and steel demand as described above. Infrastructure developments (port and rail expansions) include all projects operational and well under construction (with an operational date of 2023 at the latest). The only revision of the Wood-Mackenzie data is the production capacity of the New Tavan Tolgoi mine. The original data suggests an expansion of production to 30 Mtpa to happen as soon as the rail line to Ganqimaodu is operational, but puts that date at 2040. As the rail line is confirmed operational as of late 2021, we assume production capacity of this mine can grow to 30 Mtpa from 2022 onwards. Average production costs are increased to reflect the new production level, on the basis of the real production costs listed for 2040. 

Australian mine expansion: includes mine expansions labelled as committed or feasible in the Resources and Energy Quarterly \cite{OfficeoftheChiefEconomist-Dept.ofindustryScience2021Resources2021}, with the same production costs assumed for the expansions as for the original mines. Newly built mine projects are ignored as we consider these unlikely, and have no feasible way of estimating production costs form these.

Mongolian mine expansion: includes a further expansion of the New Tavan Tolgoi mine. Current operational capacity, with all coal trucked out, is 15 Mtpa. In this scenario we expand production capacity to 45 Mtpa, on the basis that 30 Mtpa may be transported via rail, and another 15 Mtpa via truck.

Linear coking coal price premia: with coking coal price premia following a linear relationship with CSR number. Coal market analysts indicated to us that global and Asian markets have a different appreciation for hard coking coals, with prices in the Asian market following a more linear relation with CSR.

Steel moved inland: a scenario where a greater share of steel production is moved toward inland provinces. The current split between inland vs coastal provinces in steel production is 36.4\% vs 63.6\%. For this scenario we use a 50/50 split instead.

Steel moved to ports: a scenario that represents a government policy for greater efficiency in steel production. In this scenario, for the 5 biggest coastal producers of steel (Hebei, Jiangsu, Liaoning, Shandong and Guangdong, together good for 53.3\% of all Chinese steel production), we move 25\% of existing landlocked production capacity into existing ports.

All provinces to reduce coal-fired power gen. equally: scenario where 2025 or 2030 coal-fired power generation for each province is calculated by multiplying 2019 power generation levels with the same factor. This ignores differences in recent growth rates of coal-fired power over different provinces, and effectively puts more of the decarbonization effort on inland provinces, compared with the baseline scenario.

Richer provinces reduce coal-fired power gen. faster: scenario where national reductions in coal-fired power generation for 0225 and 2030 are related to GDP. Provinces with twice the national average GDP have to reduce power production by twice the national average percentage point reduction, etc. Effectively means richer, coastal provinces do even more of the decarbonization effort, versus the baseline scenario.

No domestic rail expansion: uses 2015 rail infrastructure for all later years.

No Haoji rail line: removes the Hoaji rail line form the rail network.

No UHV network: removes the UHV network from the transport network.

No Mongolian rail expansion: keeps rail connections into Mongolia at the level as they were in 2015.

No port expansion: keeps capacities  of key Northern coal ports (Huanghua, Qinhuangdao, Caofeidian, Jingtang, Tianjin, Rizhao, Lianyungang) at their 2015 capacity.

No infrastructure expansion: keeps all of the above mentioned infrastructure at 2015 levels.

\subsection*{Data and code availability}
Our code and data, including all files describing the different components of the network, are available via the following public repository: \href{https://github.com/JorritHimself/GT-China-coal-model}{github.com}. This repository includes a censored copy of the proprietary coal mine production and costs data, with the original data layout but with mock values, so that the code remains fully functional. We also censor the data on power and steel plants, at the request of the data provider (Global Energy Monitor), but include plant identifiers as used in these original datasets. This allows others to replicate the files whilst Global Energy Monitor retains control over distribution of their data.

\bibliography{references}

\begin{thebibliography}{10}

\bibitem{Harrahill2019FrameworkJurisdictions}
Kieran Harrahill and Owen Douglas.
\newblock {Framework development for ‘just transition’in coal producing
  jurisdictions}.
\newblock {\em Energy Policy}, 134:110990, 2019.

\bibitem{IEA/OECD2019}
{IEA/OECD}.
\newblock {IEA Coal Information Statistics}.
\newblock Technical report, IEA/OECD, Paris, 2019.

\bibitem{UNTradeStatisticsBranch2019}
{UN Trade Statistics Branch}.
\newblock {UN Comtrade database}, 2021.

\bibitem{ShanxiFenshengInformationServicesCo.Ltd.2021SXcoalData}
{Shanxi Fensheng Information Services Co. Ltd.}
\newblock {SXcoal China coal data}, 2021.

\bibitem{BP2020Statistical2019}
{BP}.
\newblock {Statistical Review of World Energy 2019}, 2020.

\bibitem{Wang2019ProvincialFairness}
Delu Wang, Kaidi Wan, Xuefeng Song, and Yun Liu.
\newblock {Provincial allocation of coal de-capacity targets in China in terms
  of cost, efficiency, and fairness}.
\newblock {\em Energy Economics}, 78:109--128, 2019.

\bibitem{WoodMackenzie2021China2021}
{Wood Mackenzie}.
\newblock {China coal supply data - version of Q1 2021}.
\newblock Technical report, Wood Mackenzie, 2021.

\bibitem{Haftendorn2012TheModel}
Clemens Haftendorn, Franziska Holz, and C~V Hirschhausen.
\newblock {The end of cheap coal? A techno-economic analysis until 2030 using
  the COALMOD-World model}.
\newblock {\em Fuel}, 102:305--325, 2012.

\bibitem{Shi2018UnintendedPolicy}
Xunpeng Shi, Bertrand Rioux, and Philipp Galkin.
\newblock {Unintended consequences of China’s coal capacity cut policy}.
\newblock {\em Energy Policy}, 113:478--486, 2018.

\bibitem{Yang2018UnifyingEnterprises}
Yongqi Yang, Ming Zeng, Song Xue, Jinyu Wang, and Yuanfei Li.
\newblock {Unifying the “dual-track” pricing mechanism for coal in China:
  Policy description, influences, and suggestions for government and generation
  enterprises}.
\newblock {\em Resources, Conservation and Recycling}, 129:402--415, 2018.

\bibitem{SPGlobalPlatts2020ChinaSources}
{S{\&}P Global Platts}.
\newblock {China adds 20 million mt to thermal coal import quotas for balance
  of 2020: sources}, 2020.

\bibitem{TheGlobalCarbonProjectGCP2021GlobalAtlas}
{The Global Carbon Project (GCP)}.
\newblock {Global Carbon Atlas}.
\newblock Technical report, The Global Carbon Project (GCP), 2021.

\bibitem{EmberClimate2020GlobalReview}
{Ember Climate}.
\newblock {Global electricity review}.
\newblock Technical report, Ember Climate, 2020.

\bibitem{WorldSteelAssociation2021World2020}
{World Steel Association}.
\newblock {World steel in figures 2020}.
\newblock Technical report, World Steel Association, 2021.

\bibitem{ChinaElectricityCouncil2020Basic2019}
{China Electricity Council}.
\newblock {Basic electricity statistics 2019}, 2020.

\bibitem{Xuan2016ForecastChina}
Yanni Xuan and Qiang Yue.
\newblock {Forecast of steel demand and the availability of depreciated steel
  scrap in China}.
\newblock {\em Resources, Conservation and Recycling}, 109:1--12, 2016.

\bibitem{Dong2021ChinasPaths}
Liang Dong, Gaoyi Miao, and Weigang Wen.
\newblock {China’s Carbon Neutrality Policy: Objectives, Impacts and Paths}.
\newblock {\em East Asian Policy}, 13(01):5--18, 2021.

\bibitem{Rioux2016}
Bertrand Rioux, Philipp Galkin, Frederic Murphy, and Axel Pierru.
\newblock {Economic impacts of debottlenecking congestion in the Chinese coal
  supply chain}.
\newblock {\em Energy Economics}, 60:387--399, 2016.

\bibitem{Paulus2011CoalMarket}
Moritz Paulus and Johannes Tr{\"{u}}by.
\newblock {Coal lumps vs. electrons: How do Chinese bulk energy transport
  decisions affect the global steam coal market?}
\newblock {\em Energy Economics}, 33(6):1127--1137, 2011.

\bibitem{TheLantauGroup2020CuttingEvolves}
{The Lantau Group}.
\newblock {Cutting the Gordian Knot: China’s High-voltage Super Grid
  Evolves}, 2020.

\bibitem{Truby2012MarketTrade}
Johannes Truby and Moritz Paulus.
\newblock {Market structure scenarios in international steam coal trade}.
\newblock {\em The Energy Journal}, 33(3), 2012.

\bibitem{Ansari2020Between2055}
Dawud Ansari and Franziska Holz.
\newblock {Between stranded assets and green transformation:
  Fossil-fuel-producing developing countries towards 2055}.
\newblock {\em World Development}, 130:104947, 2020.

\bibitem{Auger2021TheAssets}
Thomas Auger, Johannes Tr{\"{u}}by, Paul Balcombe, and Iain Staffell.
\newblock {The future of coal investment, trade, and stranded assets}.
\newblock {\em Joule}, 6 2021.

\bibitem{Jie2021TheStrategies}
Dingfei Jie, Xiangyang Xu, and Fei Guo.
\newblock {The future of coal supply in China based on non-fossil energy
  development and carbon price strategies}.
\newblock {\em Energy}, 220:119644, 2021.

\bibitem{Zhang2018ASystem}
Yaru Zhang, Tieju Ma, and Fei Guo.
\newblock {A multi-regional energy transport and structure model for China’s
  electricity system}.
\newblock {\em Energy}, 161:907--919, 2018.

\bibitem{Li2019TheMarket}
Jianglong Li, Chunping Xie, and Houyin Long.
\newblock {The roles of inter-fuel substitution and inter-market contagion in
  driving energy prices: Evidences from China’s coal market}.
\newblock {\em Energy Economics}, 84:104525, 2019.

\bibitem{Wu2020ReducingApproach}
Wei Wu and Boqiang Lin.
\newblock {Reducing overcapacity in China’s coal industry: A real option
  approach}.
\newblock {\em Computational Economics}, 55(4):1073--1093, 2020.

\bibitem{WoodMackenzie2020China2020}
{Wood Mackenzie}.
\newblock {China coal supply summary - version of June 2020}.
\newblock Technical report, Wood Mackenzie, 2020.

\bibitem{NDRC2017Notice2163}
{NDRC}.
\newblock {Notice of the NDRC on deepening the market-oriented reform of
  railway freight transport charges and other related issues. NDRC Price [2017]
  No. 2163}, 2017.

\bibitem{WoodMackenzie2020Transport2020}
{Wood Mackenzie}.
\newblock {Transport cost in China - November 2020}.
\newblock Technical report, Wood Mackenzie, 2020.

\bibitem{CCIDConsulting2020WhiteConstruction}
{CCID Consulting}.
\newblock {White Paper on UHV Industry Development and Investment Opportunities
  under the "New Infrastructure Construction"}, 2020.

\bibitem{Zhitan2020ThoughtsPlan}
Liu Zhitan.
\newblock {Thoughts on the development of China's gas-fired power generation
  industry in the "14th Five-year Plan"}, 2020.

\bibitem{BiomassObserver2020TheReleased}
{Biomass Observer}.
\newblock {The 2020 China biomass power generation industry development report
  was released}, 2020.

\bibitem{IEA2020Chemicals.Needed}
{IEA}.
\newblock {Chemicals. More efforts needed}.
\newblock Technical report, OECD/IEA, 2020.

\bibitem{Shen2016ChangesChina}
Xiangjin Shen and Binhui Liu.
\newblock {Changes in the timing, length and heating degree days of the heating
  season in central heating zone of China}.
\newblock {\em Scientific reports}, 6(1):1--10, 2016.

\bibitem{OfficeoftheChiefEconomist-Dept.ofindustryScience2021Resources2021}
Energy Office of the Chief Economist Dept.~of industry, Science and Resources.
\newblock {Resources and Energy Quarterly: March 2021}.
\newblock Technical report, DISER, 2021.

\bibitem{GlobalEnergyMonitor2021GlobalTracker}
{Global Energy Monitor}.
\newblock {Global Coal Plant Tracker}, 2021.

\bibitem{IEA2020World2020}
{IEA}.
\newblock {World Energy Outlook 2020}.
\newblock Technical report, OECD/IEA, Paris, 2020.

\bibitem{MinistryofIndustryandInformationTechnology2021GuidingIndustry}
{Ministry of Industry and Information Technology}.
\newblock {Guiding opinions on promoting the high-quality development of the
  iron and steel industry}.
\newblock Technical report, MIIT, 2021.

\end{thebibliography}

\pagebreak
\section*{Supplementary material}
% Reset page numbering 
\setcounter{page}{1}
% Reset fig and Table counter and add supplementary to figure name
\renewcommand\figurename{Supplementary Fig.} % Figure naming convention
\setcounter{figure}{0}
\renewcommand\tablename{Supplementary Table} % Table naming convention
\setcounter{table}{0}

\subsection*{Changes in coal supply by origin including domestic suppliers}
\begin{figure}[H] % hbtp for here, top, bottom, next available page. H for really just here
    \centering
        \includegraphics[width=1\textwidth]{fig/Changes_Mt_China.png}
    \caption{Sensitivity of both domestic supply and Chinese imports of thermal and coking coal to changes in domestic consumption. 
    \textmd{As changes (Mt) in 2025 or 2030 versus 2019.}}
    \label{fig:scenariosMtinclChina}
\end{figure}

\pagebreak
\subsection*{Alternative presentations of sensitivity results}
\begin{figure}[H] % hbtp for here, top, bottom, next available page. H for really just here
    \centering
        \includegraphics[width=1\textwidth]{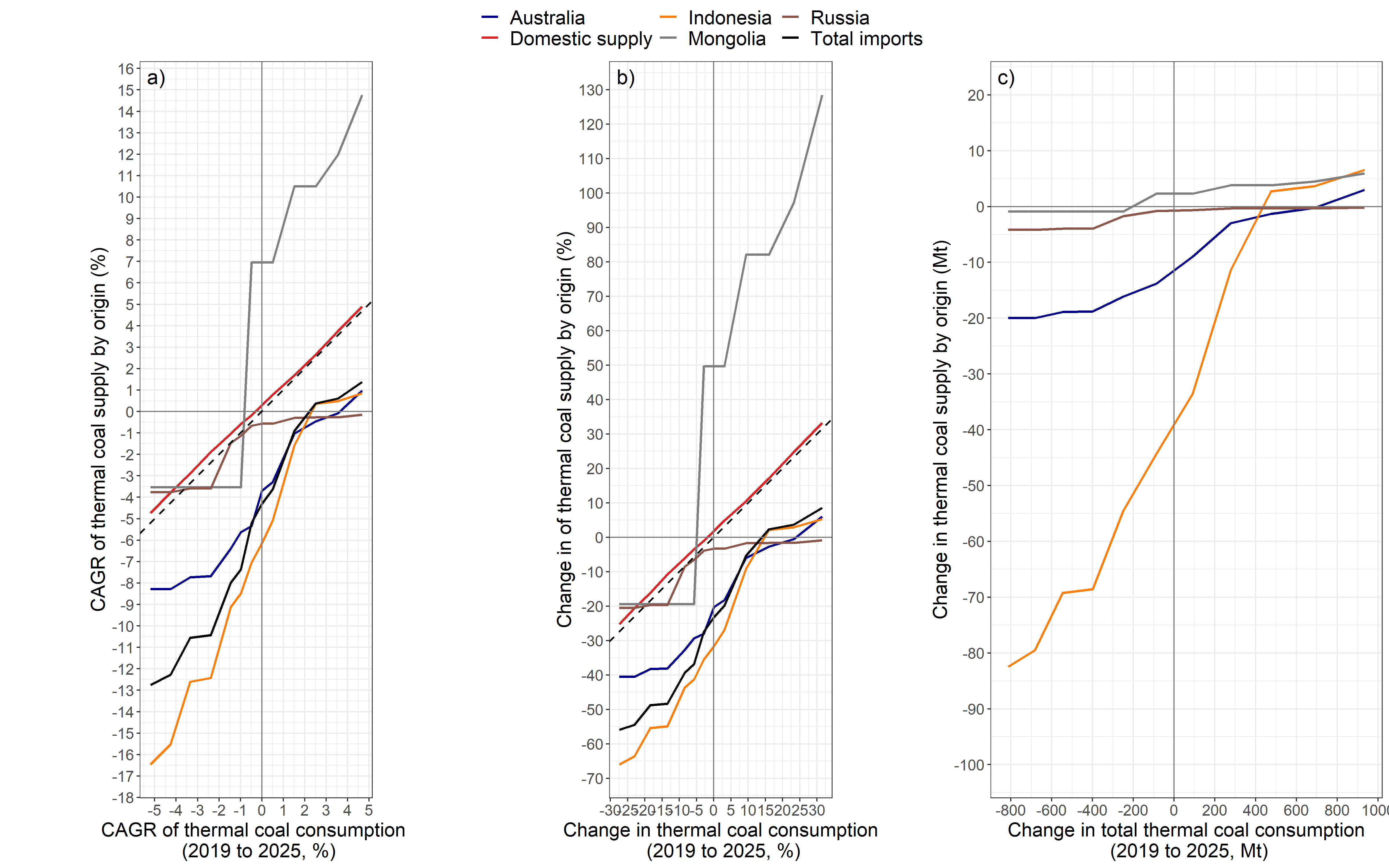}
    \caption{Sensitivity of imports to changes in Chinese consumption of thermal coal. 
    \textmd{As changes in annual growth rates (a); as total percentage growth through 2025 (b); and as Mt changes through 2025 (c).}}
    \label{fig:sensitivity_thermal_alt}
\end{figure}
\begin{figure}[H] % hbtp for here, top, bottom, next available page. H for really just here
    \centering
        \includegraphics[width=1\textwidth]{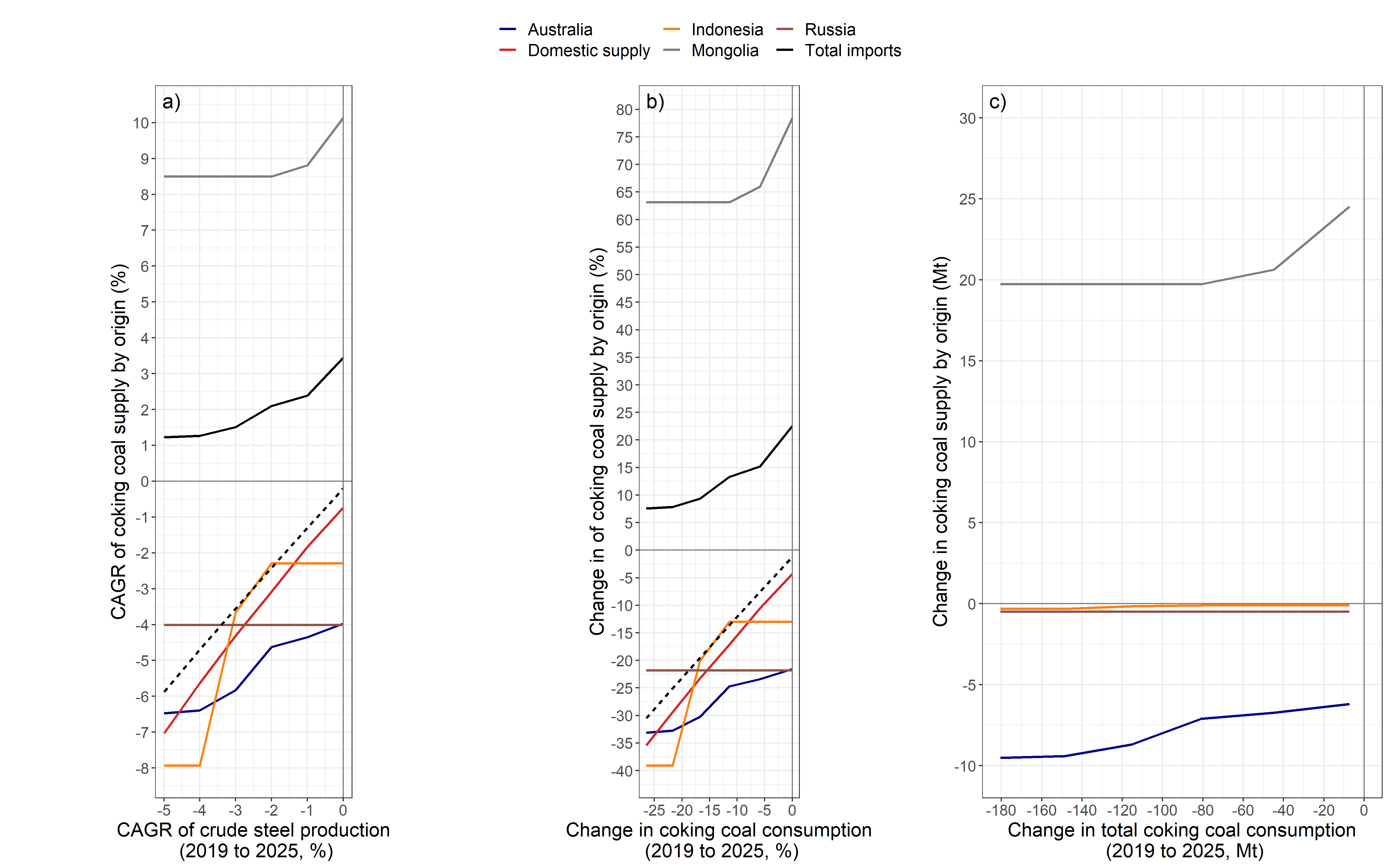}
    \caption{Sensitivity of imports to changes in Chinese consumption of coking coal. 
    \textmd{As changes in annual growth rates (a); as total percentage growth through 2025 (b); and as Mt changes through 2025 (c).}}
    \label{fig:sensitivity_coking_alt}
\end{figure}

\pagebreak
\subsection*{Alternative presentations of scenario results}
\begin{figure}[H] % hbtp for here, top, bottom, next available page. H for really just here
    \centering
        \includegraphics[width=1\textwidth]{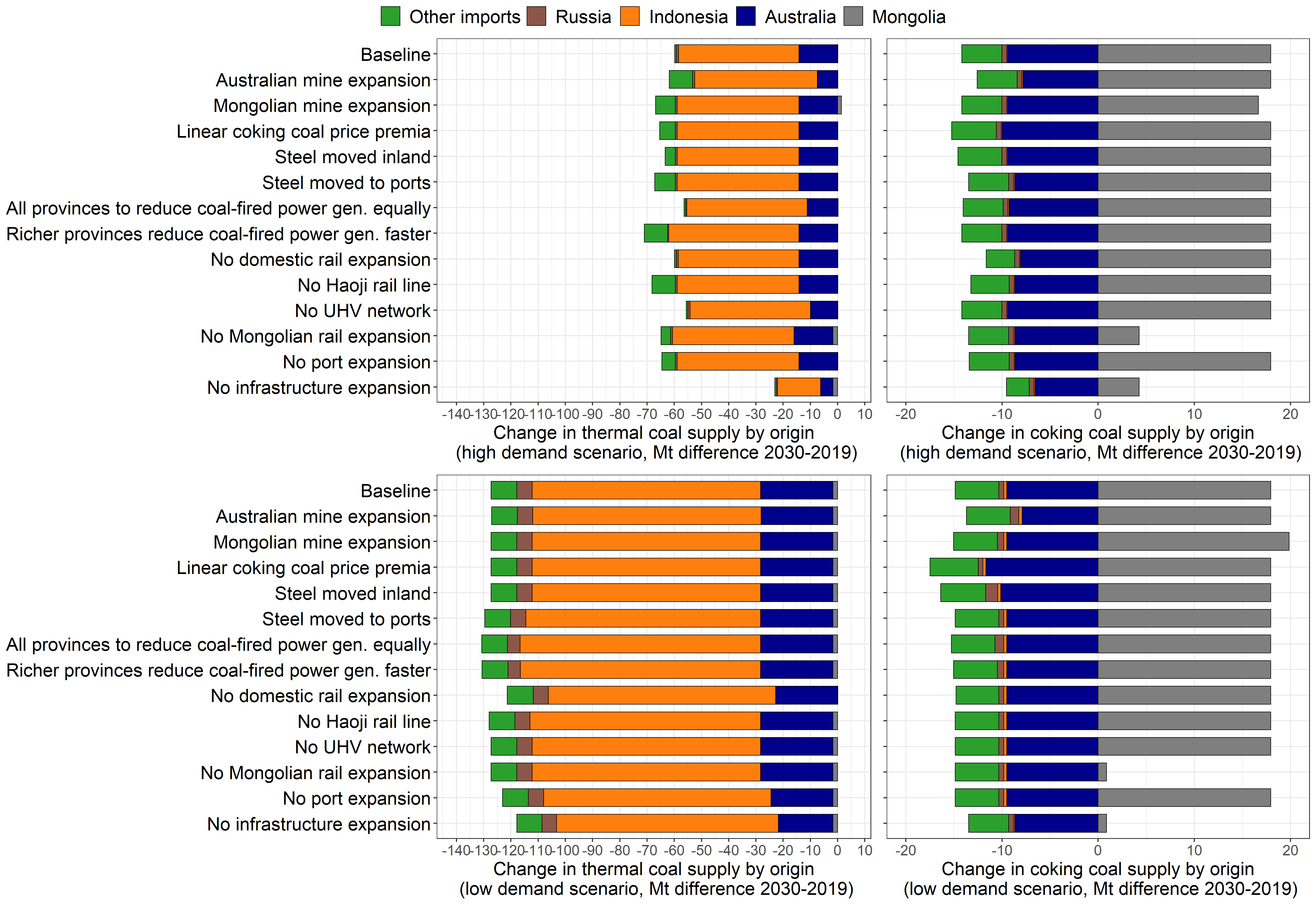}
    \caption{Scenarios for Chinese market and infrastructure development and effect on imports. 
    \textmd{As changes (Mt) in 2030 versus 2019.}}
    \label{fig:scenarios2030}
\end{figure}
\begin{figure}[H] % hbtp for here, top, bottom, next available page. H for really just here
    \centering
        \includegraphics[width=1\textwidth]{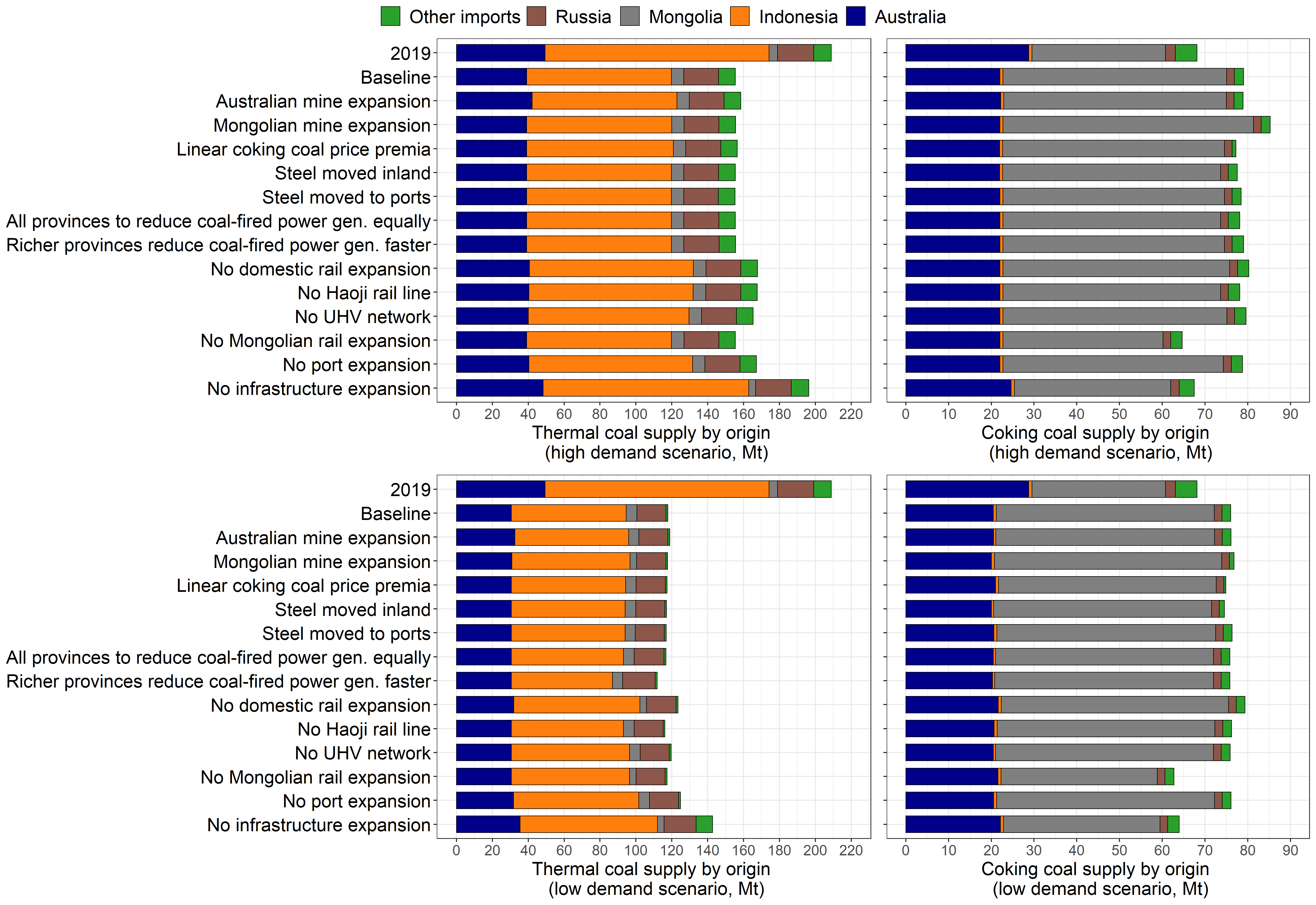}
    \caption{Scenarios for Chinese market and infrastructure development and effect on imports. 
    \textmd{As total imports (Mt) by 2025.}}
    \label{fig:scenariosMt2025}
\end{figure}
\begin{figure}[H] % hbtp for here, top, bottom, next available page. H for really just here
    \centering
        \includegraphics[width=1\textwidth]{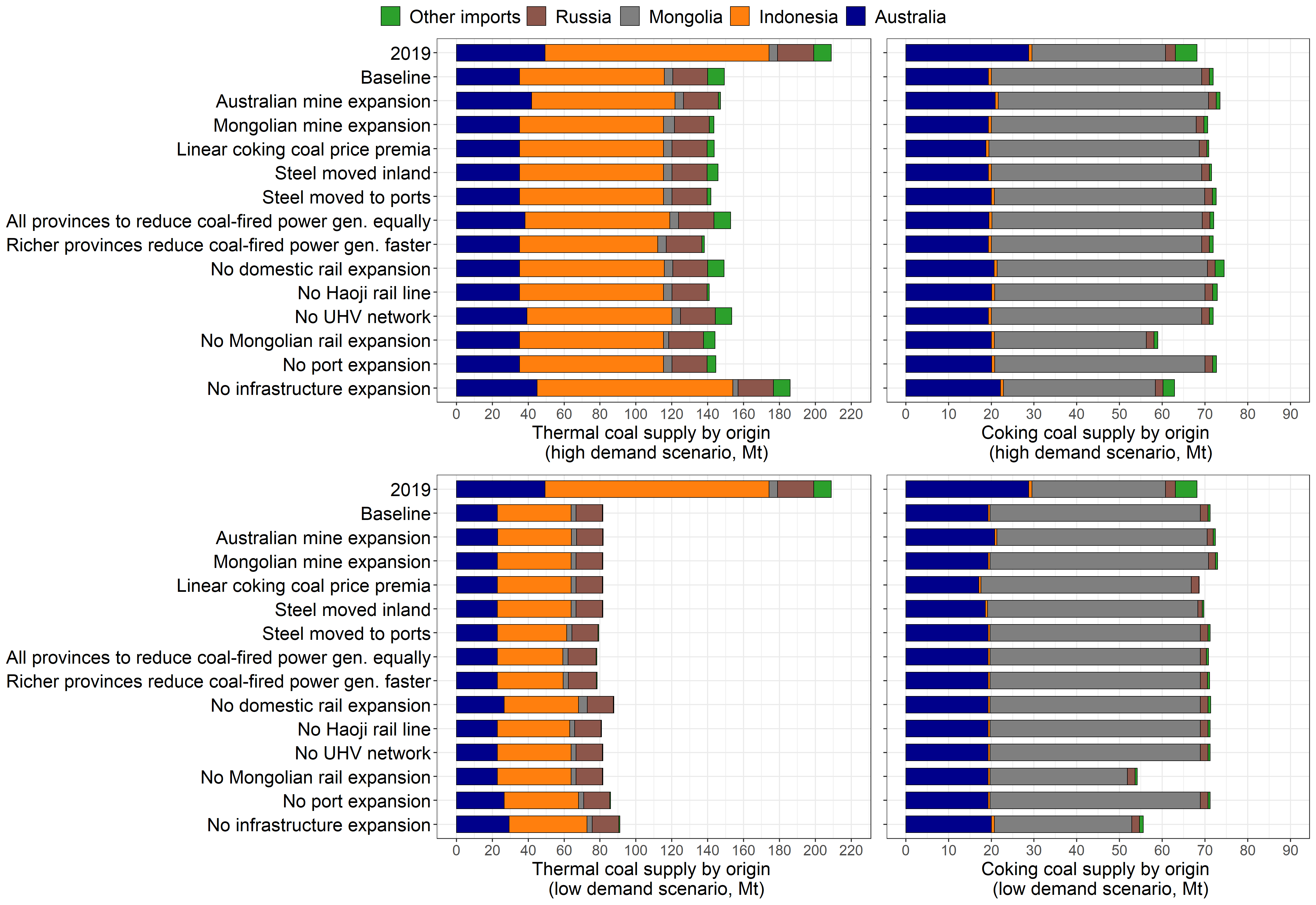}
    \caption{Scenarios for Chinese market and infrastructure development and effect on imports. 
    \textmd{As total imports (Mt) by 2030.}}
    \label{fig:scenariosMt2030}
\end{figure}

\pagebreak
\subsection*{Chinese coal mine capacity}
\begin{figure}[H] % hbtp for here, top, bottom, next available page. H for really just here
    \centering
        \includegraphics[width=1\textwidth]{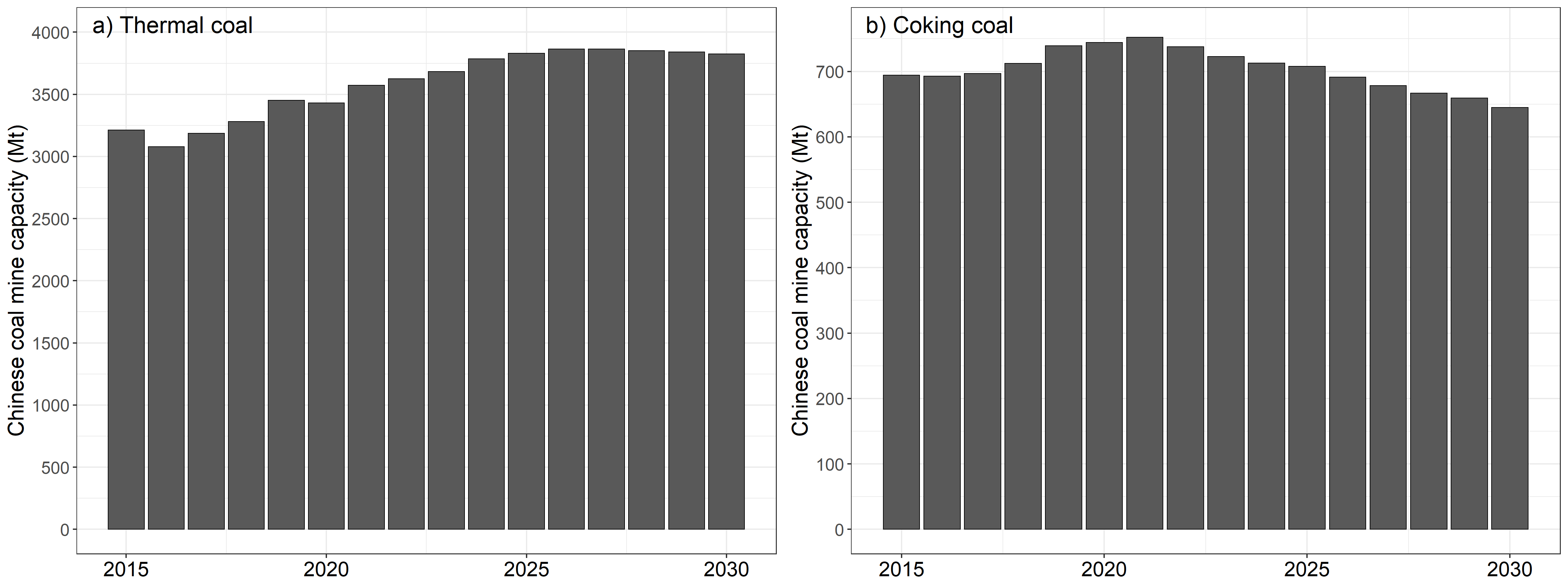}
    \caption{Chinese coal mining capacity.
    \textmd{Note: as physical Mt. Data from Wood MacKenzie's baseline view for China \cite{WoodMackenzie2020China2020}.}}
    \label{fig:mine_capa}
\end{figure}

\pagebreak
\subsection*{Global supply curve correction}
Global production curves are adjusted to account for the fact that Chinese customers will not be able to buy all of the cheapest coals. We consider the key suppliers to China, and determine the shares of exports to China in their total exports (average share over 2017-2019), and multiply the production capacity of each mine in those key supplier countries with that same percentage. 

To illustrate, consider that a certain country's supply curve looks like panel (a) in Supplementary Fig. \ref{fig:glob_supply_curves}. If we know that 20\% of this country's exports over the period 2017-2019 were destined for China, we multiply each of the supply steps, i.e., the technical capacity of each individual mine, with that same 20\%; panel (b) in Supplementary Fig. \ref{fig:glob_supply_curves}. That results in a corrected supply curve of that country, as Chinese consumers experience it, as in panel (c) in Supplementary Fig. \ref{fig:glob_supply_curves}. 

We apply these correction factors to each of the exports of the six countries that supply more than 97\% of China's imports; all other countries are grouped under `Rest of world'.
The corrections applied are calculated by dividing Chinese imports data from SXcoal\cite{ShanxiFenshengInformationServicesCo.Ltd.2021SXcoalData} with total export data from the IEA\cite{IEA/OECD2019}. The correction factors applied, separately for thermal and coking coal, are listed in Supplementary Table \ref{tab:glob_supply_curves}.

\begin{figure}[H] % hbtp for here, top, bottom, next available page. H for really just here
    \centering
        \includegraphics[width=1\textwidth]{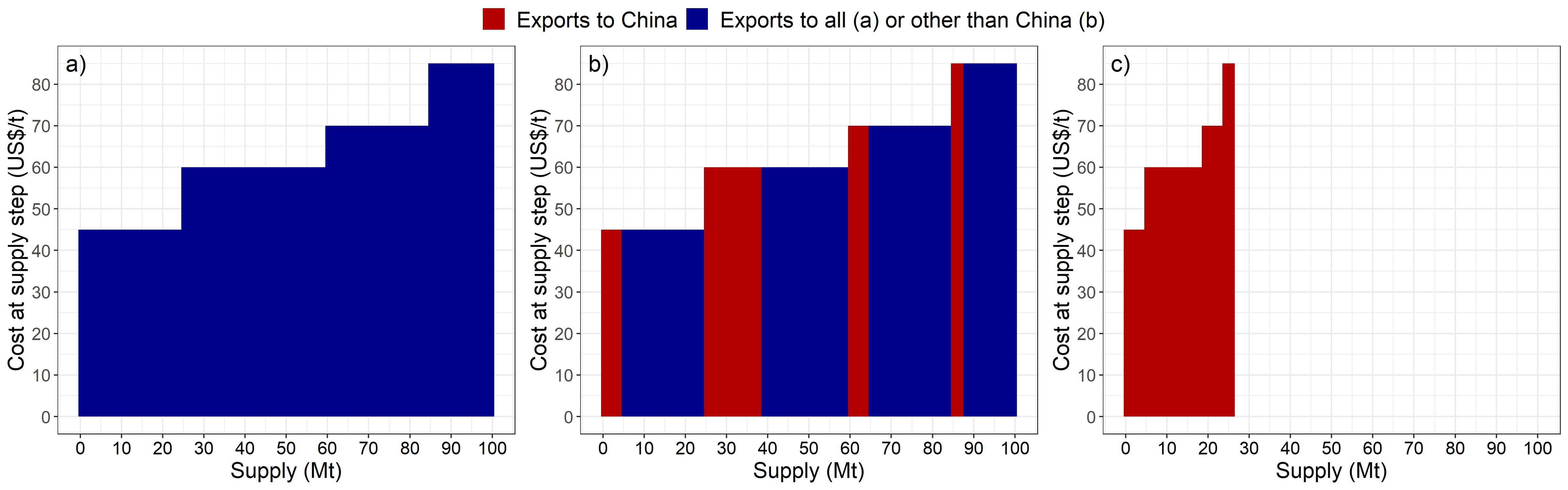}
    \caption{Illustration of our correction of global supply curves. 
    \textmd{}}
    \label{fig:glob_supply_curves}
\end{figure}

\begin{table}[H]
\centering
\caption{Correction factors for global supply curves}
\label{tab:glob_supply_curves}
\begin{tabular}{lll}
\toprule
Country of origin & Thermal coal & Coking coal \\ 
\midrule
Australia         & 24.1\%       & 16.8\%      \\
Indonesia         & 29.3\%       & 8.1\%       \\
Russia            & 14.3\%       & 19.6\%      \\
Mongolia          & 100.0\%      & 100.0\%     \\
Canada            & 55.8\%       & 9.9\%       \\
Philippines       & 76.3\%       & n/a          \\
Rest of world     & 1.4\%        & 3.1\%       \\
\midrule
\end{tabular}
\end{table}

\pagebreak
\subsection*{Coal qualities and price premia}
We use Wood MacKenzie's coal supply data on coal prices and qualities to determine price premia, with price data for 2019.

For thermal coals, we found a relation between calorific value and price of \$15 per 1000 kcal/kg (Price excl. premium in Supplementary Fig. \ref{fig:premiathermal}), plus a price premium of \$8 per 1000 kcal/kg. versus a benchmark 5500 kcal/kg coal (Price incl. premium Supplementary Fig. \ref{fig:premiathermal}).

For coking coals, we group coals with different Coking Strength after Reaction (CSR no.) and calculate average price within each group. This gives a non-linear price premium of \$40/t for the highest quality coking coal; see panel (a) in Supplementary Fig. \ref{fig:premiacoking}. Coal market analysts from Wood MacKenzie informed us they would typically assume a perfectly linear relation between CSR no. and prices in the Asian coking coal market, as graphed in panel (b) in Supplementary Fig. \ref{fig:premiacoking}. 
We apply the price premia as graphed in panel (b) in our scenario `linear coking coal premia'; all our other scenarios use the values as graphed in panel (a).

\begin{figure}[H] % hbtp for here, top, bottom, next available page. H for really just here
    \centering
        \includegraphics[width=0.7\textwidth]{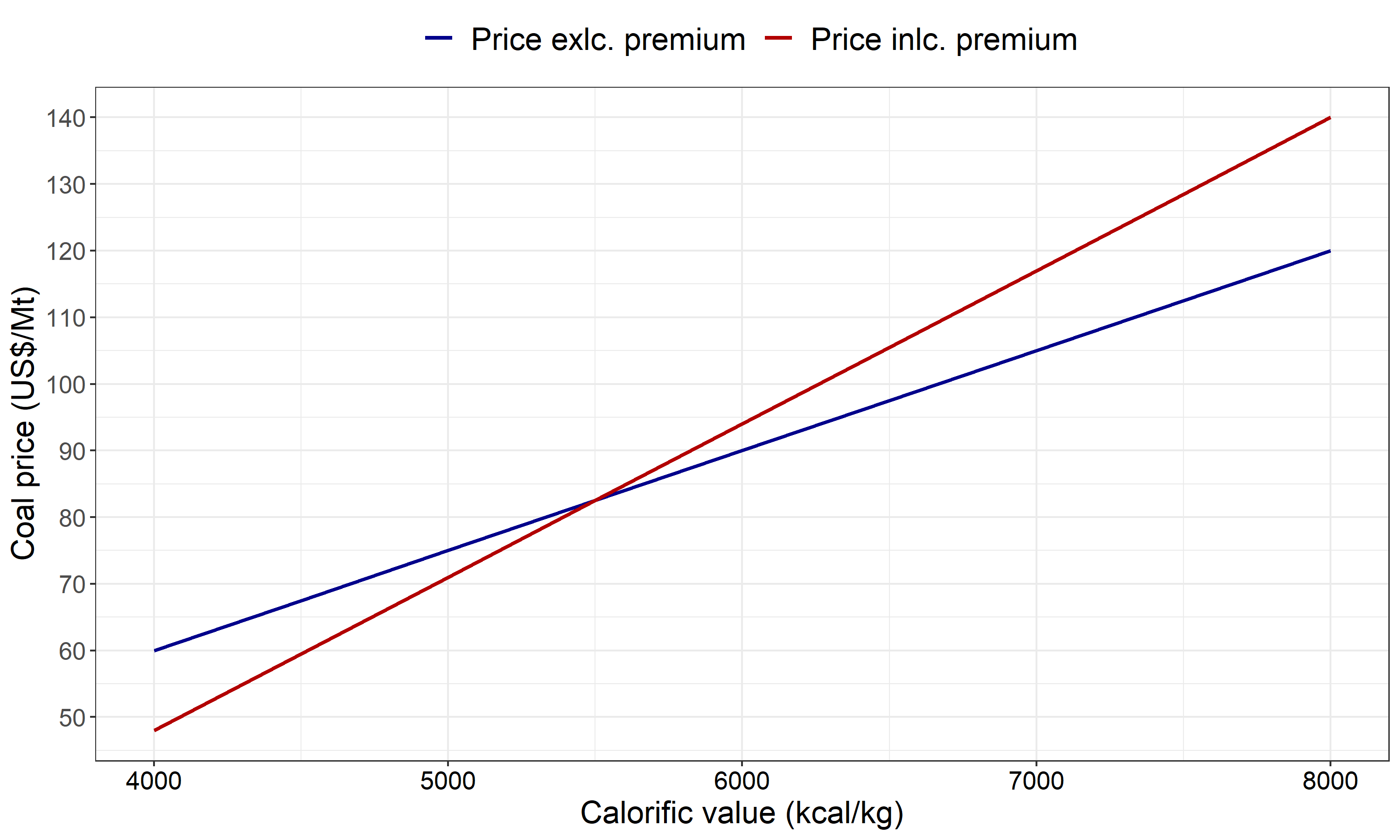}
    \caption{Assumed price premia for thermal coals in the Chinese market. 
    \textmd{}}
    \label{fig:premiathermal}
\end{figure}

\begin{figure}[H] % hbtp for here, top, bottom, next available page. H for really just here
    \centering
        \includegraphics[width=1\textwidth]{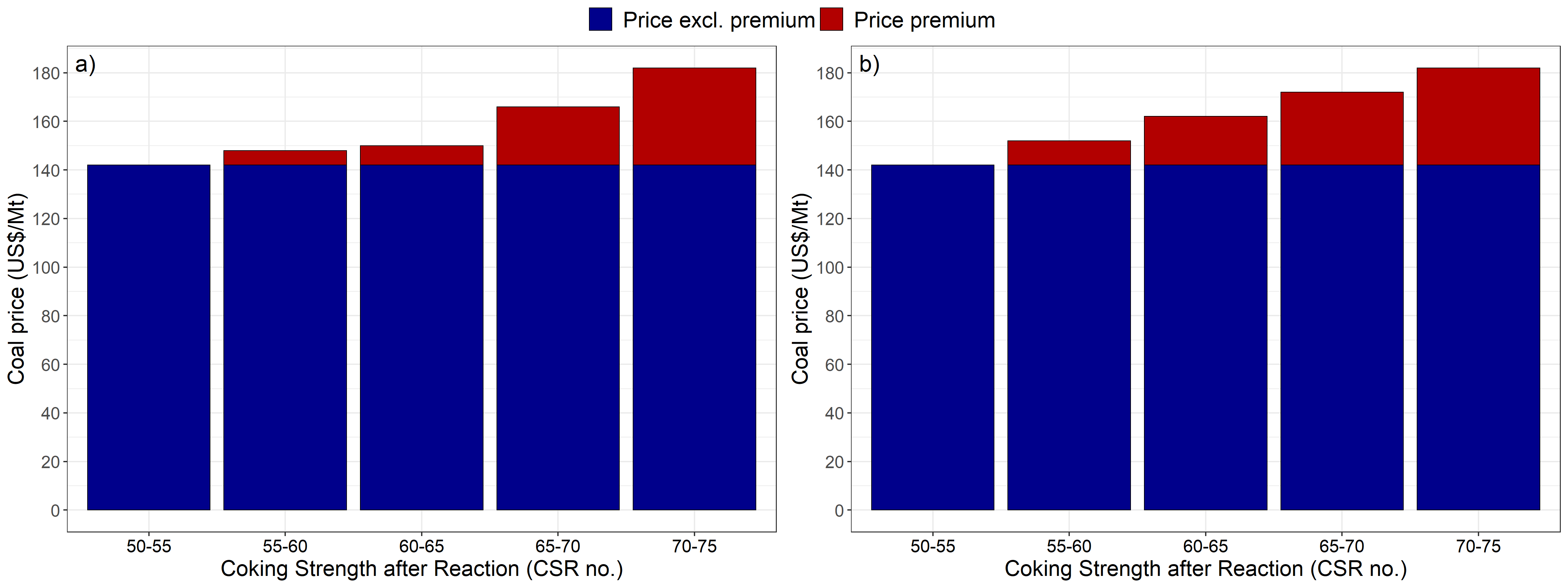}
    \caption{Assumed price premium paid for high quality coking coals in the Chinese market. 
    \textmd{}}
    \label{fig:premiacoking}
\end{figure}

\pagebreak
\subsection*{Assumed mix of coking coal for primary steel making}
\begin{figure}[H] % hbtp for here, top, bottom, next available page. H for really just here
    \centering
        \includegraphics[width=0.7\textwidth]{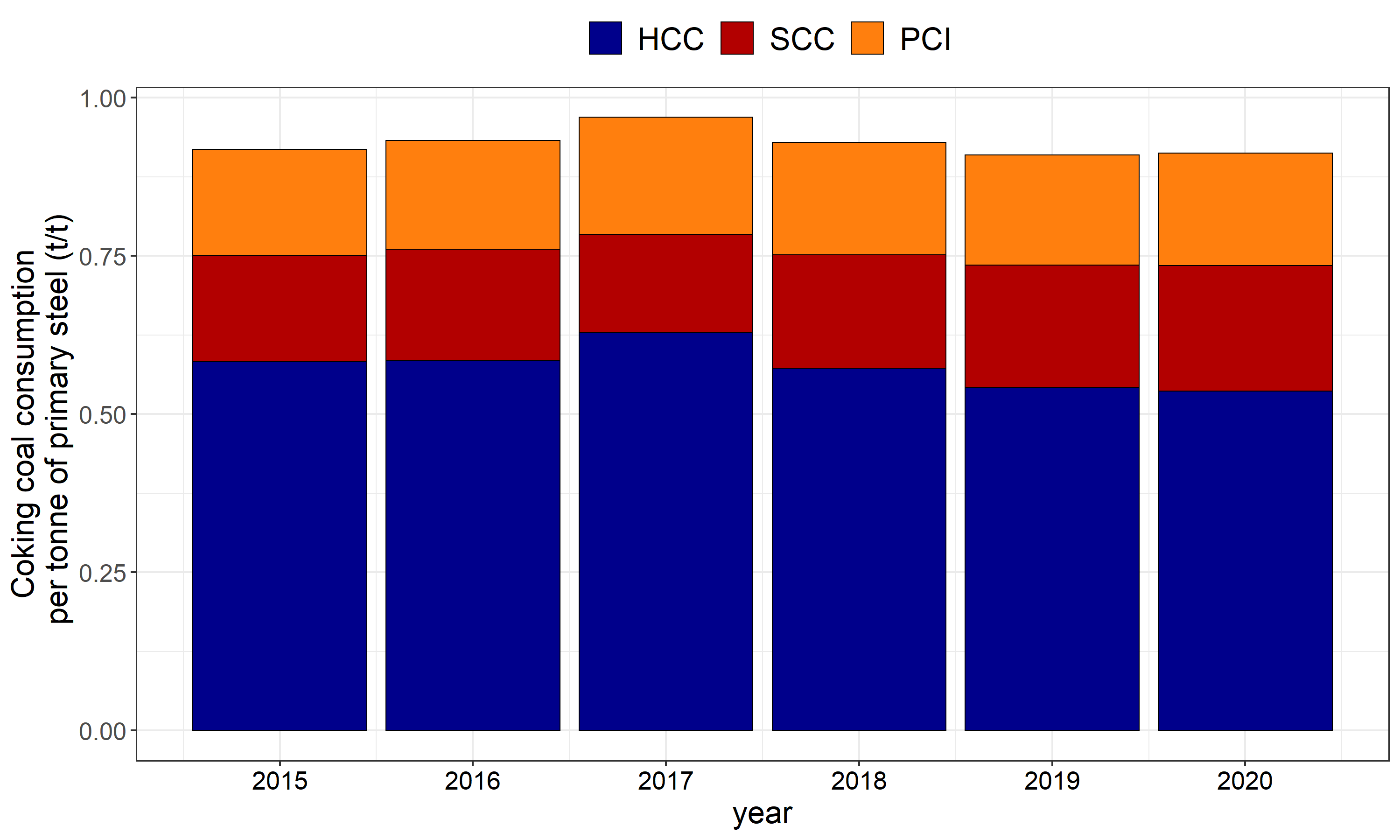}
    \caption{Mix of coking coal consumed in China for every tonne of of primary steel produced. 
    \textmd{Note: HCC = Hard Coking Coal, SCC = Semi-soft Coking Coal, PCI = Pulverized coal for Injection.}}
    \label{fig:mix_coking_coal}
\end{figure}

To derive a mix of coking coal consumed for primary steel production, we use domestic and Mongolian coking coal supply data from Wood Mackenzie\cite{WoodMackenzie2021China2021}, which splits this out in three different types of coking coal (HCC, SCC, and PCI), and add coking coal import data from SXcoal\cite{ShanxiFenshengInformationServicesCo.Ltd.2021SXcoalData}. We use crude steel production data from the World Steel Association \cite{WorldSteelAssociation2021World2020}, and subtract scrap consumption data provided by a China steel market analyst at Morgan Stanley. We are not terribly convinced of data quality for the pandemic year of 2020, and so we average coking coal consumption per tonne of primary steel over 2017-2019, resulting in a mix of 581 kg HCC, 176 kg SCC, and 179 kg PCI (Supplementary Fig. \ref{fig:mix_coking_coal}). 

\pagebreak
\subsection*{National level coal energy balance adjustment}
To determine the total amount of energy from coal consumed, and how it is distributed over the country, we start from SXcoal's national level data on coal consumption by industry (Supplementary Table \ref{tab:balance_orig_Mt}).

We convert these physical Mt to energy content using the calorific values provided by SXcoal (Supplementary Table \ref{tab:balance_orig_PJ}). 

From this overview, we've removed all thermal and coking coal consumed for steel making. The estimated mix of coking coals from our own estimation method, matches that volume nearly exactly (Supplementary Table \ref{tab:coking_calibration}). We therefore consider our mix of Hard coking coal (HCC), Semi-soft Coking Coal (SCC) and Pulverized coal for Injection (PCI) to correspond with what SXcoal reports as thermal and coking coal consumed in the steel making sector. This is not unexpected; other statistical data sources on coal also tend to classify either Semi-soft Coking Coal or Pulverized coal for Injection as either thermal or coking coal. We simply refer to all three types of coal used in steel making as coking coal. 

We then recalculate SXcoal's provincial level data on coal-fired power generation, and use the fleet-wide capacity weighted average conversion efficiency to estimate the total amount of coal consumed as primary energy input (Supplementary Table \ref{tab:powergen_calibration}). The implied primary coal consumption from these statistics is roughly 15\% higher than the number for coal consumption by the power generation industry in in SXcoal's national level statistics on coal consumption by industry (Supplementary Table \ref{tab:balance_orig_PJ}). 

We therefor substitute the number derived from provincial-level statistics on coal-fired power generation, and reduce the coal consumption under `other uses' (Supplementary Table \ref{tab:balance_adj_PJ}) such that the total level of coal consumption is equal with SXcoal's original total reported in Supplementary Table \ref{tab:balance_orig_PJ}.

\begin{table}[hbtp]
\caption{China's coal consumption by industry (originally reported, as Mt)}
\centering
\begin{threeparttable}
\begin{tabular}{llllllll}
\toprule
\textbf{}                      & \textbf{2015} & \textbf{2016} & \textbf{2017} & \textbf{2018} & \textbf{2019} & \textbf{2020} & CV   \\
\midrule
Power generation               & 1,757         & 1,819         & 1,881         & 2,011         & 2,058         & 2,085         & 4,700 \\
Building Materials             & 311           & 315           & 314           & 288           & 321           & 322           & 5,000 \\
Chemicals                      & 161           & 162           & 179           & 175           & 194           & 199           & 5,500 \\
Iron \& Steel (thermal coal)   & 136           & 139           & 144           & 154           & 160           & 176           & 6,000 \\
Heat Supply                    & 198           & 220           & 232           & 265           & 287           & 289           & 4,700 \\
Others                         & 636           & 439           & 398           & 375           & 362           & 369           & 5,000 \\
\midrule
Total thermal coal consumption & 3,199         & 3,094         & 3,148         & 3,268         & 3,381         & 3,441         &      \\
Iron \& Steel (coking coal)    & 528           & 534           & 515           & 520           & 552           & 558           &      \\
\midrule
Total coal consumption         & 3,727         & 3,628         & 3,662         & 3,788         & 3,933         & 3,999         &     \\
\midrule
\end{tabular}
    \begin{tablenotes}
      \small
      \item Note: consumption as physical Mt. The CV values are industry wide averages. Data source: SXcoal\cite{ShanxiFenshengInformationServicesCo.Ltd.2021SXcoalData}.
    \end{tablenotes}
\end{threeparttable}
\label{tab:balance_orig_Mt}
\end{table}

\begin{table}[hbtp]
\caption{China's coal consumption by industry (converted to PJ)}
\centering
\begin{threeparttable}
\begin{tabular}{lllllll}
\toprule
\textbf{}                      & \textbf{2015} & \textbf{2016} & \textbf{2017} & \textbf{2018} & \textbf{2019} & \textbf{2020} \\
\midrule
Power generation               & 34,548        & 35,770        & 36,988        & 39,543        & 40,467        & 41,011        \\
Building Materials             & 6,516         & 6,595         & 6,573         & 6,021         & 6,708         & 6,743         \\
Chemicals                      & 3,694         & 3,726         & 4,120         & 4,023         & 4,474         & 4,585         \\
Iron \& Steel (thermal)        & \multicolumn{6}{c}{Determined separately; see Supplementary Table \ref{tab:coking_calibration}}\\
Heat Supply                    & 3,902         & 4,329         & 4,553         & 5,215         & 5,636         & 5,691         \\
Others                         & 13,298        & 9,189         & 8,318         & 7,854         & 7,568         & 7,712         \\
\midrule
Total thermal coal consumption & 61,957        & 59,609        & 60,552        & 62,656        & 64,852        & 65,742        \\
Iron \& Steel (coking)         & \multicolumn{6}{c}{Determined separately; see Supplementary Table \ref{tab:coking_calibration}}      \\   
\midrule
\end{tabular}
\end{threeparttable}
\label{tab:balance_orig_PJ}
\end{table}

\begin{table}[hbtp]
\caption{Coking coal calibration}
\centering
\begin{threeparttable}
\begin{tabular}{lllllll}
\toprule
\textbf{}                                                      & \textbf{2015} & \textbf{2016} & \textbf{2017} & \textbf{2018} & \textbf{2019} & \textbf{2020} \\
\midrule
Primary steel production (Mt)                                  & 719.6         & 721.5         & 679.8         & 724.9         & 782.5         & 804.8         \\
HCC consumption (Mt, our estimate)                             & 418.1         & 419.2         & 395.0         & 421.1         & 454.6         & 467.6         \\
SCC consumption (Mt, our estimate)                             & 126.6         & 127.0         & 119.7         & 127.6         & 137.7         & 141.6         \\
PCI consumption (Mt, our estimate)                             & 128.8         & 129.1         & 121.7         & 129.7         & 140.1         & 144.1         \\
\midrule
Total coking coal consumption (Mt, our   estimate)             & 673.5         & 675.3         & 636.3         & 678.5         & 732.4         & 753.2         \\
Total coking + thermal coal cons. for   steel (SXcoal) & 664           & 673           & 659           & 674           & 712           & 733           \\
\midrule
Difference our method vs SXcoal data                           & 10            & 3             & -22           & 5             & 21            & 20           \\
\midrule
\end{tabular}
    \begin{tablenotes}
      \small
      \item Note: we estimate consumption of coking coal at an average mix of 581 kg of HCC, 176 kg of SCC, and 179 kg of PCI, per tonne of primary steel production, see Supplementary Fig. \ref{fig:mix_coking_coal}.
    \end{tablenotes}
\end{threeparttable}
\label{tab:coking_calibration}
\end{table}

\begin{table}[hbtp]
\caption{Coal-fired power generation calibration}
\centering
\begin{threeparttable}
\begin{tabular}{>{\raggedright}p{6cm}llllll}
\toprule
\textbf{} & \textbf{2015} & \textbf{2016} & \textbf{2017} & \textbf{2018} & \textbf{2019} & \textbf{2020} \\
\midrule
Coal-fired power production (TWh) & 3,849     & 4,029     & 4,215     & 4,547     & 4,694     & 4,765     \\
Coal-fired power production (PJ)  & 13,855        & 14,503        & 15,172        & 16,371        & 16,898        & 17,153        \\
Fleet wide avg. efficiency        & 34.56\%       & 34.94\%       & 35.21\%       & 35.49\%       & 35.69\%       & 35.96\%       \\
\midrule
Implied primary coal consumption (PJ)     & 40,093        & 41,508        & 43,090        & 46,128        & 47,346        & 47,695        \\
Difference with national-level stats by industry (PJ)     & 5,545        & 5,738        & 6,102        & 6,56        & 6,879        & 6,684        \\
\midrule
\end{tabular}
    \begin{tablenotes}
      \small
      \item Note: coal-fired power production data from SXcoal\cite{ShanxiFenshengInformationServicesCo.Ltd.2021SXcoalData}; fleet-wide average conversion efficiency data from Global Energy Monitor\cite{GlobalEnergyMonitor2021GlobalTracker}.
    \end{tablenotes}
\end{threeparttable}
\label{tab:powergen_calibration}
\end{table}

\begin{table}[hbtp]
\caption{China's coal consumption by industry (adjusted, PJ)}
\centering
\begin{threeparttable}
\begin{tabular}{lllllll}
\toprule
\textbf{}                      & \textbf{2015} & \textbf{2016} & \textbf{2017} & \textbf{2018} & \textbf{2019} & \textbf{2020} \\
\midrule
Power generation               & 40,093        & 41,508        & 43,090        & 46,128        & 47,346        & 47,695        \\
Building Materials             & 6,516         & 6,595         & 6,573         & 6,021         & 6,708         & 6,743         \\
Chemicals                      & 3,694         & 3,726         & 4,120         & 4,023         & 4,474         & 4,585         \\
Iron \& Steel                  & \multicolumn{6}{c}{Determined separately; see Supplementary Table \ref{tab:coking_calibration}}\\
Heat Supply                    & 3,902         & 4,329         & 4,553         & 5,215         & 5,636         & 5,691         \\
Others                         & 7,753         & 3,450         & 2,216         & 1,269         & 689           & 1,028         \\
\midrule
Total thermal coal consumption & 61,957        & 59,609        & 60,552        & 62,656        & 64,852        & 65,742       
       \\
\midrule
\end{tabular}
\end{threeparttable}
\label{tab:balance_adj_PJ}
\end{table}

\clearpage
\subsection*{Coal consumption forecasts}
For our comparisons in Fig. \ref{fig:scenarios}, we use both a low and high demand growth scenario. These are based on the IEA's `Stated Policies Scenario' and its `Sustainable Development Scenario'. These have previously been shown to represent the upper and lower range of a suite of demand forecasts from different organizations, in a review by Auger et al\cite{Auger2021TheAssets}.

We take the numbers on coal demand from the IEA's World Energy Outlook 2020\cite{IEA2020World2020}. The Annex A to this report provides forecasts for 2025, and 2030, including separate forecasts for the power generation sector. It does not provide separate forecasts for steel making. For the development of coal consumption in the high demand scenario, we use a forecast provided to us by a China steel market analyst at Morgan Stanley. This forecast presumes a high and extended post-Covid construction stimulus, with total steel demand at 1015 Mt in 2025 and 922 Mt in 2030, with a supply of steel from scrap to grow to 290 Mt in 2025, and 322 Mt in 2030, up from 214 Mt in 2019. For the low demand scenario, we use the `2 degree compatible' forecast by Wood MacKenzie, which sees total steel demand of 985 Mt in 2025 and 882 Mt in 2030, and use a goal for steel from scrap of 330 Mt by 2025 and 400 Mt by 2030, taken from a recent policy document by China's Ministry of Industry and Information Technology\cite{MinistryofIndustryandInformationTechnology2021GuidingIndustry}. We include these numbers in the IEA's scenario and thus derive implied growth rates for `Other', which is the growth rates we use for all other industries apart from power generation and steel making (Supplementary Tables \ref{tab:IEA_SPS}\&\ref{tab:IEA_SDS}).

\begin{table}[hbtp]
\caption{Forecasted Chinese coal consumption in the high demand Scenario}
\centering
\begin{threeparttable}
\begin{tabular}{p{1.6cm}p{1.2cm}p{1.2cm}p{1.2cm}p{1.7cm}p{1.7cm}p{2cm}p{2cm}}
\toprule
                 &       &       &       & CAGR      & CAGR      & Total change & Total change \\
                 & 2019  & 2025  & 2030  & 2019-2025 & 2019-2030 & 2019-2025    & 2019-2030    \\
\midrule
Power            & 2,263 & 2,348 & 2,324 & 0.6\%     & 0.2\%     & 3.8\%        & 2.7\%        \\
Steel            & 732   & 679   & 562   & -1.3\%    & -2.4\%    & -7.3\%       & -23.2\%      \\
Other            & 837   & 822   & 832   & -0.3\%    & -0.1\%    & -1.8\%       & -0.6\%       \\
\midrule
Total            & 3,832 & 3,849 & 3,717 & 0.1\%     & -0.3\%    & 0.4\%        & -3.0\%        \\
\midrule
\end{tabular}
    \begin{tablenotes}
      \small
      \item Note: based on the IEA's `Stated Policies Scenario' (SPS). For `Power' and 'Other': thermal coal as Mt of coal with an average calorific value of 5,000 kcal/kg coal; for `Steel': as physical Mt of coking coal with a mix as determined in Supplementary Fig. \ref{fig:mix_coking_coal}.
    \end{tablenotes}
\end{threeparttable}
\label{tab:IEA_SPS}
\end{table}

\begin{table}[hbtp]
\caption{Forecasted Chinese coal consumption in the low demand scenario}
\centering
\begin{threeparttable}
\begin{tabular}{p{1.6cm}p{1.2cm}p{1.2cm}p{1.2cm}p{1.7cm}p{1.7cm}p{2cm}p{2cm}}
\toprule
      &       &       &       & CAGR      & CAGR      & Total change & Total change \\
      & 2019  & 2025  & 2030  & 2019-2025 & 2019-2030 & 2019-2025    & 2019-2030    \\
\midrule
Power & 2,263 & 2,001 & 1,481 & -2.0\%    & -3.8\%    & -11.6\%      & -34.6\%      \\
Steel & 732   & 612   & 481   & -2.9\%    & -3.8\%    & -16.4\%      & -34.3\%      \\
Other & 837   & 783   & 649   & -1.1\%    & -2.3\%    & -6.5\%       & -22.4\%      \\
\midrule
Total & 3,832 & 3,396 & 2,611 & -2.0\%    & -3.4\%    & -11.4\%      & -31.9\%     \\
\midrule
\end{tabular}
    \begin{tablenotes}
      \small
      \item Note: based on the IEA's `Sustainable Development Scenario' (SDS). For `Power' and `Other': thermal coal as Mt of coal with an average calorific value of 5,000 kcal/kg coal; for `Steel': as physical Mt of coking coal with a mix as determined in Supplementary Fig. \ref{fig:mix_coking_coal}.
      \end{tablenotes}
\end{threeparttable}
\label{tab:IEA_SDS}
\end{table}

\pagebreak
\subsection*{Mechanism for distributing forecasted consumption}
\begin{figure}[H] % hbtp for here, top, bottom, next available page. H for really just here
    \centering
        \includegraphics[width=0.9\textwidth]{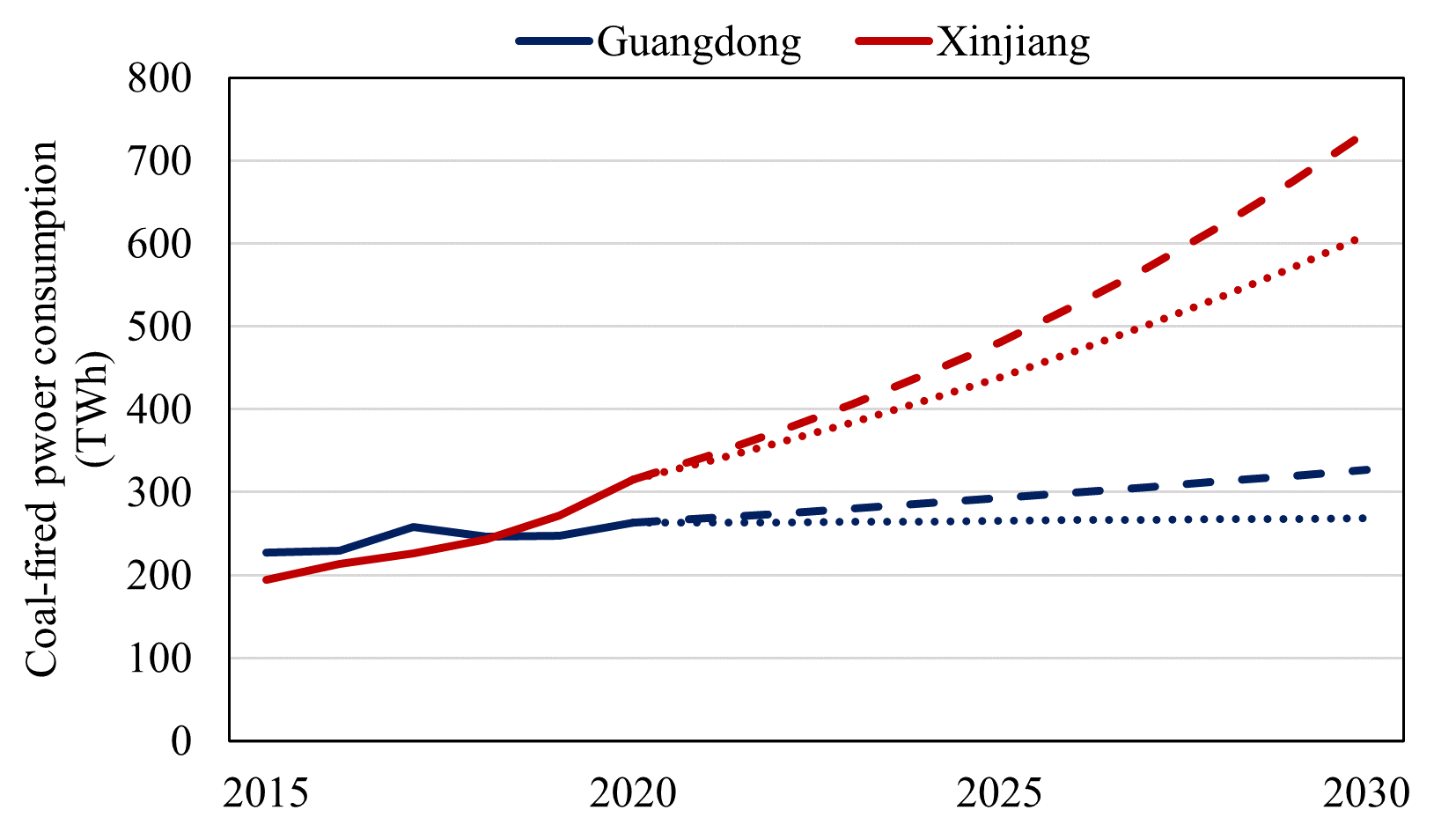}\\
    \caption{Mechanism for adjustment of provincial level forecasts of consumption levels.
    \textmd{Dashed lines are extrapolations based on historical annual growth rates, dotted lines are extrapolations after subtracting 2\% points from this hitsorical growth rates for either province.}
    }
    \label{fig:prov-forecast-adj}
\end{figure}

Different provinces have seen strongly differing growth rates of coal consumption over recent years, largely due to strongly differing levels of development, economic growth, industrial activity, and power consumption per capita, etc. See for example the development of coal-fired power consumption in Guangdong, one of China's richest, coastal provinces, and Xinjiang, a relatively poor inland province (Supplementary Fig.\ref{fig:prov-forecast-adj}). Data in this figure for 2015-2020 is historical data. 
We calculate the average annual percentage growth rate of coal-fired power generation over the period 2015 to 2019 for each province, and use this data to extrapolate through to 2030 (the dashed lines in Supplementary Fig.\ref{fig:prov-forecast-adj}). 
In order to end up at a certain national level consumption growth for 2025 or 2030, we then add or subtract a certain number of percentage points to each provinces' historical growth rates to extrapolate through to 2025 or 2030, such that sum of provincial level consumption adds up to the required national level of consumption in our scenario setting (the dotted lines in Supplementary Fig.\ref{fig:prov-forecast-adj}). 
In this example, historical growth rates were 2.2\% pa for Guangdong, and 8.8\% for Xinjiang. Both have their forecasted growth reduced by 2 percentage points, so down to 0.2\% pa for Guangdong, and 6.8\% for Xinjiang. We consider this a sensible way to reflect the notion that China's decarbonization policies will almost certainly have a distributional aspect that considers differing development levels and recent growth rates.

\subsection*{Larger versions of the maps of the transport networks in our model}
\begin{figure}[H] % hbtp for here, top, bottom, next available page. H for really just here
    \centering
        \includegraphics[angle=90,origin=c, width=0.9\textwidth]{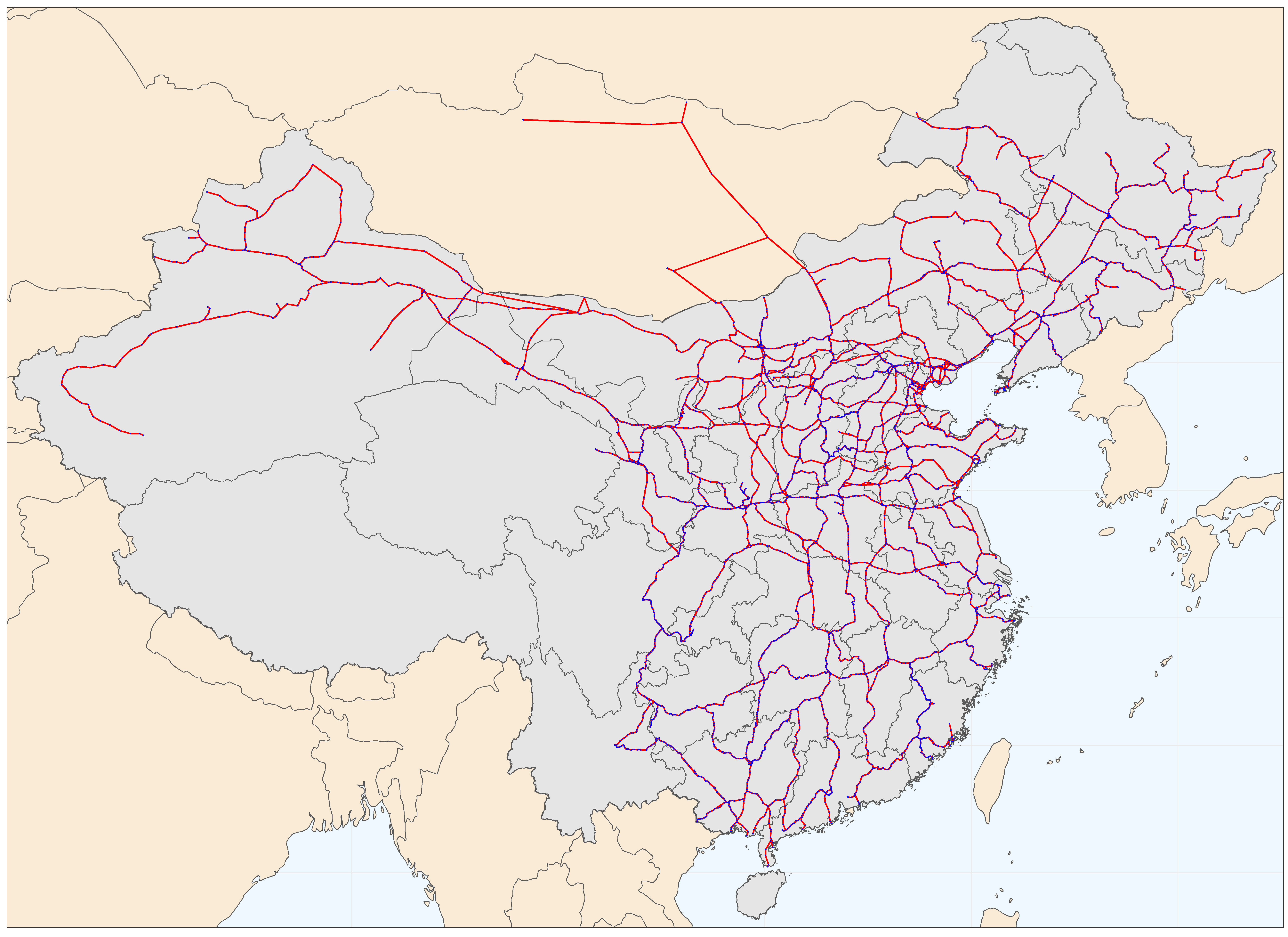} \\
    \caption{Railway network in our model.
    \textmd{Includes only freight railway lines that are used for the transport of coal; railway stops in blue, railway links in red. Main data source: China Railway Map (\url{cnrail.geogv.org}).}}
    \label{fig:rail-network}
\end{figure}
\begin{figure}[H] % hbtp for here, top, bottom, next available page. H for really just here
    \centering
        \includegraphics[angle=90,origin=c, width=1\textwidth]{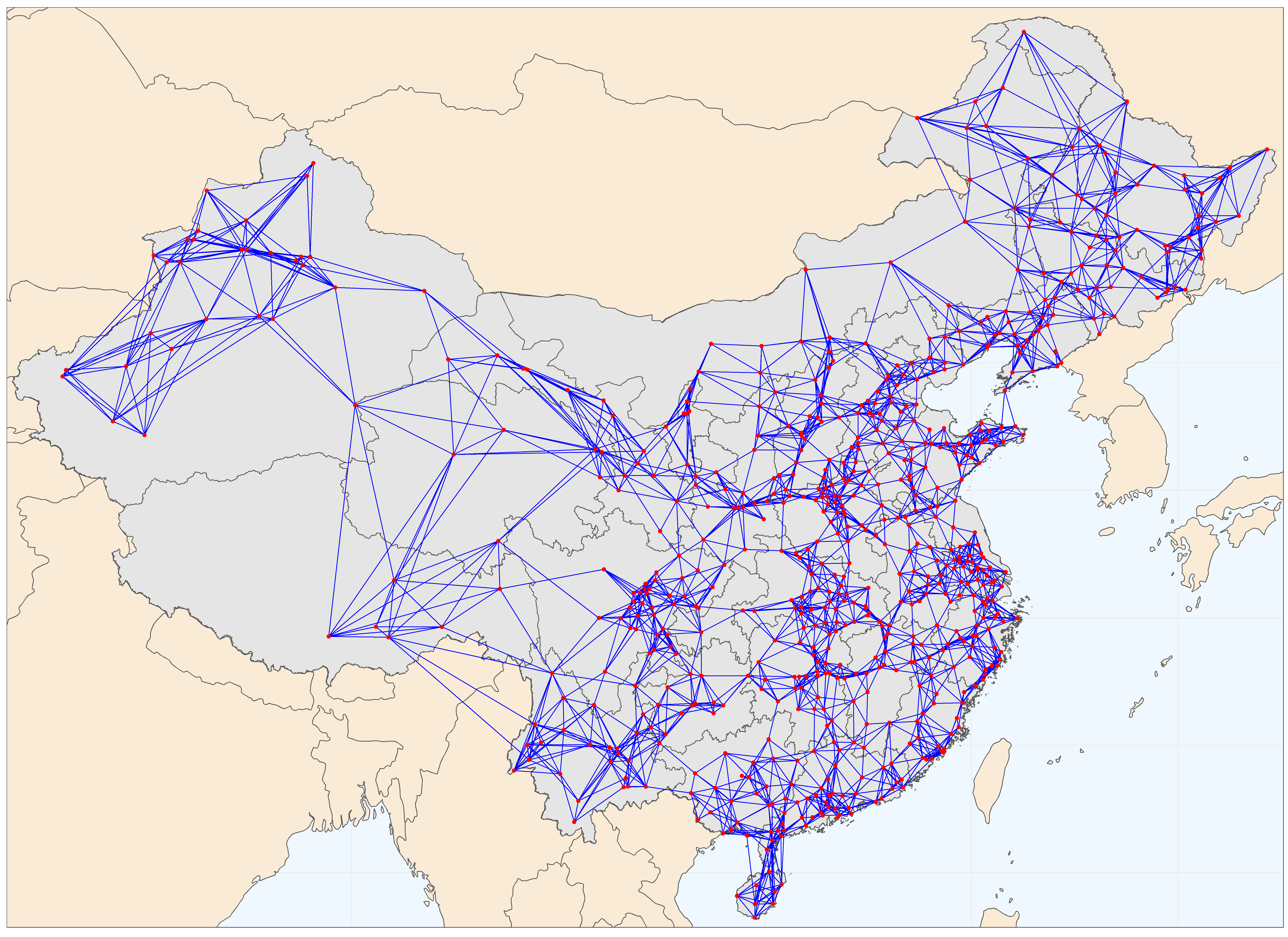} \\
    \caption{Road network in our model.
    \textmd{Inter-city road transport network; city nodes in red, road links in blue; note the links are drawn here as straight lines but actual driving distances have been used to calculate transport costs over each link.}}
    \label{fig:road-network}
\end{figure}
\begin{figure}[H] % hbtp for here, top, bottom, next available page. H for really just here
    \centering
        \includegraphics[angle=90,origin=c, width=1\textwidth]{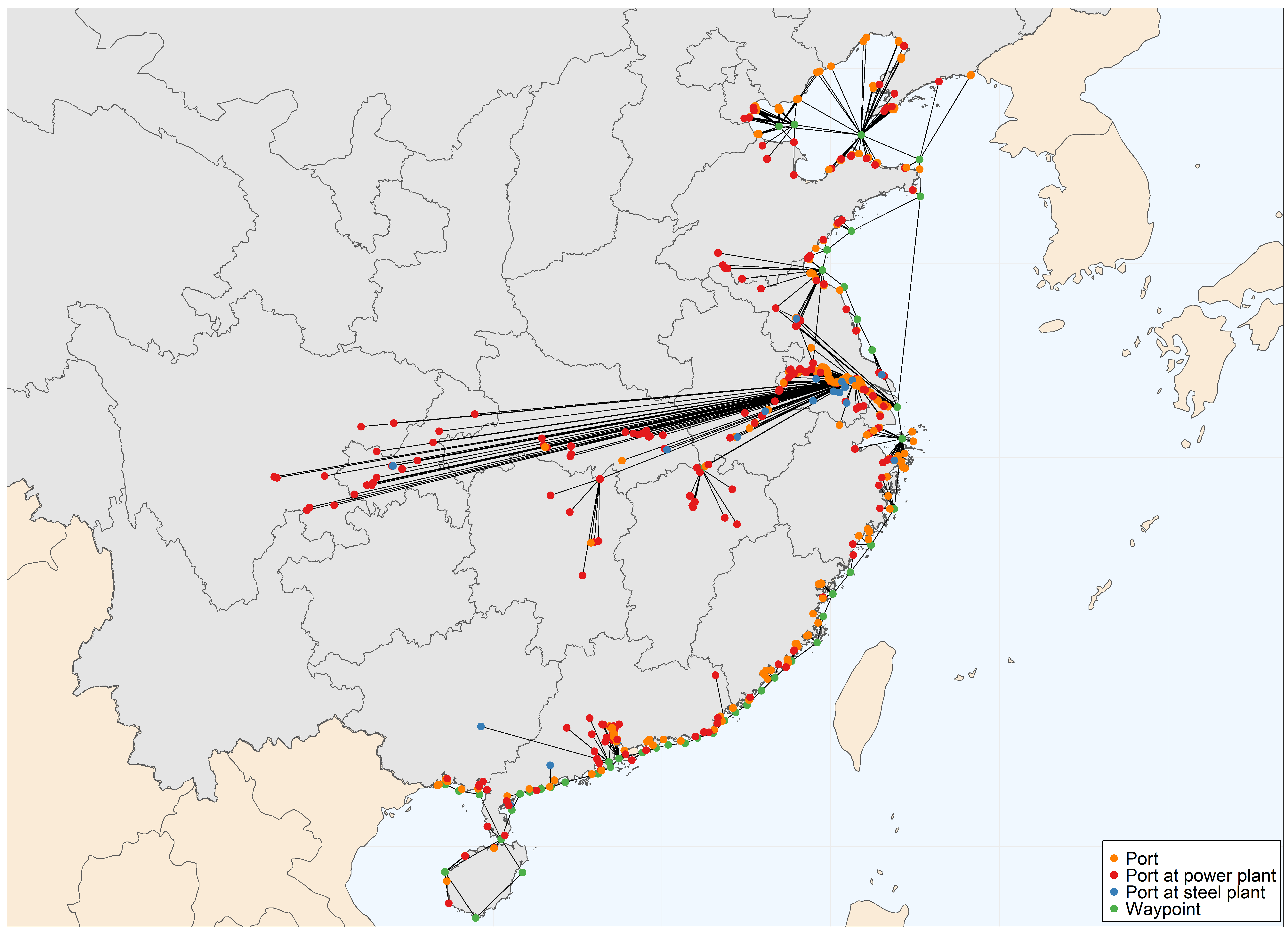} \\
    \caption{Navigation network in our model.
    \textmd{Includes ports, ports facilities at individual power and steel plants, and navigational waypoints that are used to connect the network. Main data source: Kpler (\url{kpler.com}) database for dry bulk goods.}}
    \label{fig:navi-network}
\end{figure}
\begin{figure}[H] % hbtp for here, top, bottom, next available page. H for really just here
    \centering
        \includegraphics[angle=90,origin=c, width=1\textwidth]{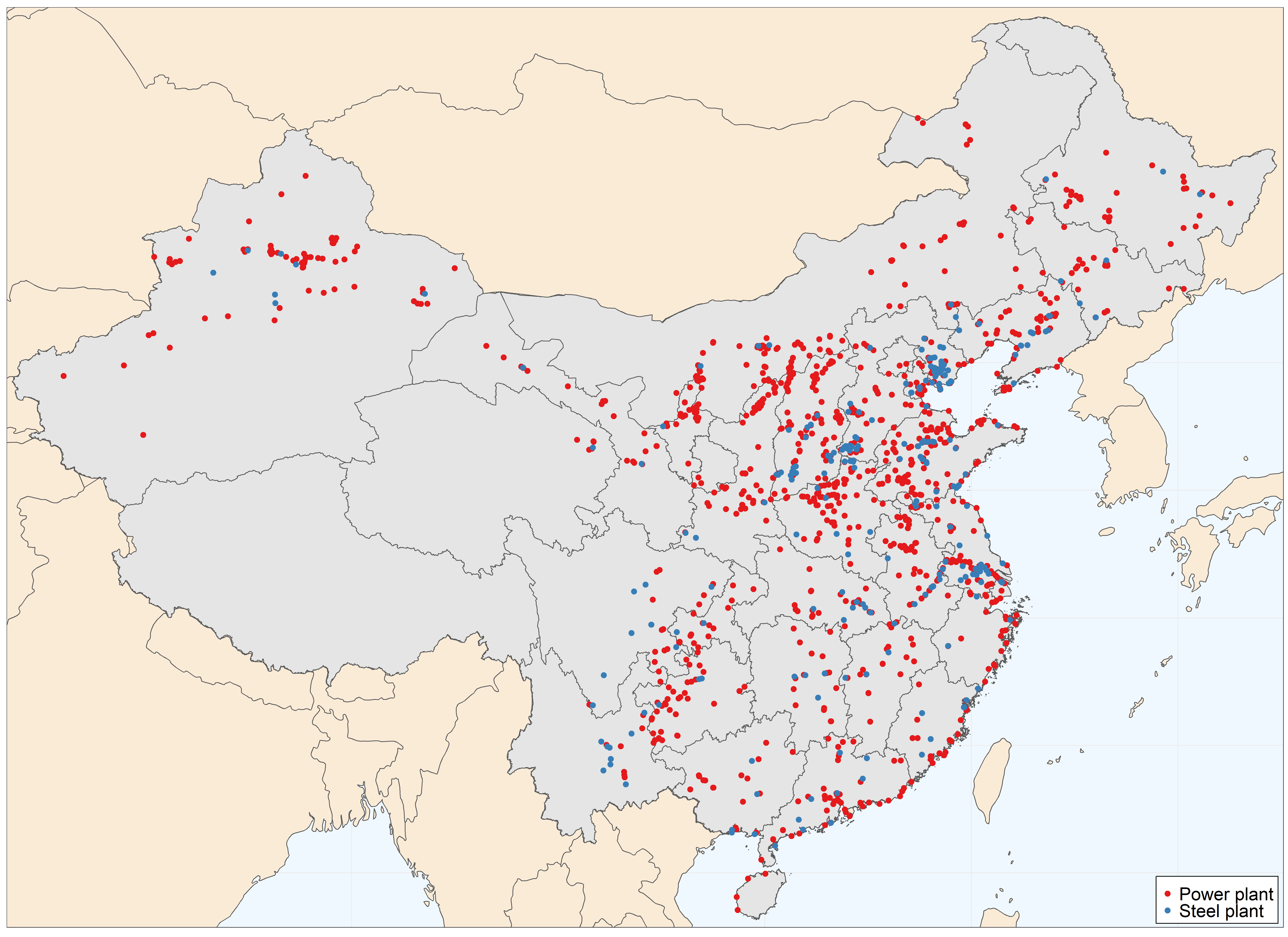} \\
    \caption{Power and steel plants in our model.
    \textmd{Main data source: global power plant tracker and the global steel plant tracker developed by Global Energy Monitor (\url{globalenergymonitor.org}).}}
    \label{fig:power-steel-plants-map}
\end{figure}

\end{document}